\documentclass[twocolumn,twoside,showpacs,amsmath,amssymb,byrevtex,altaffilletter,citeautoscript,jasa]{revtex4asa}

\usepackage{graphicx}
\usepackage{dcolumn} 
\usepackage{bm}      
\usepackage{helvet}  

\usepackage{ucs}
\usepackage[utf8x]{inputenc}

\begin{document}
        \long\def\ignore#1\endignore{}
        \def\endignore{}
        \def\eqref#1{(\ref{#1})}
        \def\sign{\,\mathrm{sign}}
        \def\ph#1{\phantom{#1}}
        \def\e#1{{\cdot}10^{#1}}
        \def\arg{\sl}
        \def\note#1{{\bf[#1]}}
        \newdimen{\width}
        \width=\columnwidth
        \def\psdir{eps}
        \def\dimvar{\hat}
        \def\althat{\tilde}

\title{Contribution to harmonic balance calculations of periodic
  oscillation for self-sustained musical instruments with focus on
  single-reed instruments}

\author{Snorre Farner}
 \affiliation{Department of Electronics and
Telecommunications, NTNU, O.~S. Bragstads pl.~2, 7491 Trondheim, Norway}%
\author{Christophe Vergez and Jean Kergomard and Aude Liz\'ee}%
 \affiliation{Laboratoire de M\'ecanique et d'Acoustique, CNRS UPR 7051, 
   31 chemin Joseph Aiguier, 13402 Marseille Cedex 20, France}%

\defineheaderauthor{S. Farner et al.}
\definerunningtitle{Harmonic balance calculations for
musical instruments}
\date{\today} 

\begin{abstract}
  
  The harmonic balance method (HBM) was originally developed for
  finding periodic solutions of electronical and mechanical systems
  under a periodic force, but has later been adapted to self-sustained
  musical instruments.  Unlike time-domain methods, this
  frequency-domain method does not capture transients and so is not
  adapted for sound synthesis. However, its independence of time makes
  it very useful for studying every periodic solution of the model,
  whether stable or unstable without care of initial conditions.  A
  computer program for solving general problems involving nonlinearly
  coupled exciter and resonator, ``Harmbal'', has been developed based
  on the HBM.  The method as well as convergence improvements and
  continuations facilities are thorougly presented and discussed in
  the present paper.  Application of the method is
  demonstrated on various problems related to a common model of
  the clarinet: a reed modelled as a simple spring with and
  without mass and damping, a nonlinear coupling and a cubic
  simplification of it, and a cylindrical bore with or without
  dissipation and dispersion as well as a bore formed as a stepped
  cone.

\end{abstract}

\maketitle
\section{Introduction}

Since Helmholtz \cite{helmholtz77}, it has become natural to describe
a self-sustained\footnote{Self-sustained is a term indicating
  oscillation driven by a constant energy input.} musical instrument as
an exciter
coupled to a resonator.  More recently, McIntyre et al.\ 
\cite{mcintyre83} have highlighted that simple models are able to
describe the main functioning of most self-sustained musical
instruments. These models rely on few equations whose
  implementation is not \textsc{cpu}-demanding, mainly because the
  nonlinearity is spatially localized in an area small compared to the
  wavelength.  This makes them well adapted for real-time computation
(including both transient and steady states). These models are
particularly popular in the framework of sound synthesis.

On the other hand, calculation in the frequency domain is suitable
for determining periodic solutions of the model (the
values of the harmonics as well as the playing frequency) for a given
set of parameters. Such information can be provided by an iterative
method named the harmonic balance method (HBM). Though the name
``harmonic balance'' seems to date back to 1936 \cite{krylov36}, the
method was popularized nearly forty years ago for electrical and
mechanical engineering purposes, first for forced vibrations
\cite{urabe65}, later for auto-oscillating systems \cite{stokes72}.
The modern version was presented rather shortly after by Nakhla and
Vlach \cite{nakhla76}.  In 1978, Schumacher was the first one to use
the HBM for musical acoustics purposes with a focus on the clarinet
\cite{schumacher78}. However in this paper, the playing frequency is
not determined by the HBM. This shortcoming is the major
improvement brought by Gilbert et al.\ \cite{gilbert89} eleven
  years later, who proposed a full study of the clarinet including
the playing frequency as an unknown of the problem.

The fact that the HBM can only calculate periodic solutions,
may seem as a drawback. Certainly, transients such as the attack are
impossible to calculate, and the periodic result is boring to listen
to and does not represent the musicality of the instrument. Therefore
the HBM is definitely not intended for sound synthesis.  Nevertheless,
self-sustained musical instruments are usually used to generate
harmonic sounds, which are periodic by definition.  The HBM
is thus very useful to investigate the behavior of a physical model
of an instrument, depending on its parameter values. This is possible
for both stable and unstable solutions, without care of precise
initial conditions. Moreover, HBM results can be compared to
approximate analytical calculations (like the variable truncation
method (VTM) \cite{kergomard00}), in order to check the validity of
the approximate model considered.

The present paper is based on the work of Gilbert et al \cite{gilbert89}.  
Our main contributions are: 
extension of the diversity of equations managed, improved convergence
of the method, introduction of basic continuation facilities, and from
a practical point of view, faster calculations.

While the main idea is already described by Gilbert et al.\
\cite{gilbert89},  
Section~\ref{s:nummeth} details the principle of the HBM, in
particular the discretization of the problem, both in time and frequency.

Section~\ref{s:harmbal} is devoted to the various contributions of
the current work, which are applied in a computer program called
Harmbal \cite{harmbal}. The framework is defined to include models with three
equations: two linear differential equations, written in the frequency
domain, and a nonlinear coupling equation in the time domain (see
Sec.~\ref{s:self-sustained}). As usual in the HBM, this system of
three equations is solved iteratively. The solving method chosen
(Newton-Raphson, Sec.~\ref{s:harmdet}) has been
investigated and its convergence has been improved through a
backtracking scheme (Secs.~\ref{s:holes}
and~\ref{s:backtracking}). 

To illustrate the advantages of the HBM and the improvements, a few
case studies were performed and are presented in Section~\ref{s:case}.
They are based on a classical model of single reed instruments which
is presented in Section~\ref{s:clarinet}. In Sections~\ref{s:verif}
and further, simplifications to each of the three equations are
introduced so that the results could be compared to analytical
calculations, both for cylindrical and stepped-cones bores. Finally
the full model is compared to time-domain simulations.  This also
shows the modularity of Harmbal.  The comparison is achieved through
the investigation of bifurcation diagrams as the dimensionless blowing
pressure is altered. The derivation of a branch of solution is
obtained thanks to basic continuation with an
auto-adaptative parameter 
step.

Finally, various questions are tackled through practical experience
from using Harmbal. Section~\ref{s:practexp} discusses multiplicity of
solutions and poor robustness in the frequency estimation.

\section{Numerical method}\label{s:nummeth}

\subsection{The harmonic balance method}\label{s:HBM}

The harmonic balance method is a numerical method to calculate the
steady-state spectrum of periodic solutions of a
nonlinear dynamical system.  In this paper we are only concerned with
periodic solutions.  The following provides a detailed and general
description of the method for a nonlinearly coupled exciter-resonator
system.

Let $X(\omega_k)$, $k = 0, \dots, N_t-1$ be the Discrete
Fourier transform (DFT) of one period $x(t)$, $0\le t<T$, of a
$T$-periodic solution of a mathematical system to be defined.
$X(\omega_k)$ will have a number of complex components $N_t$, which
depends on the sampling frequency $f_s=1/T_s$ with which we discretize
$x(t)$ into $N_t=T/T_s$ equidistant samples.  Furthermore,
$\omega_k{=}2\pi f_p T_s k$ is the angular frequency of each harmonic of the
fundamental frequency $f_p$ of the oscillation, referred to as the {\em playing
frequency}. Note that the sampling frequency $f_s=N_tf_p$ is automatically
adjusted to the current playing frequency so that we always
consider one period of the oscillation while keeping $N_t$
constant. Note also that $N_t$ should be sufficiently large to avoid
aliasing. Moreover, if it is chosen a power of two, the Fast Fourier
transform (FFT) may be used. Assuming that $N_p<N_t/2$ harmonics
is sufficient to 
describe the solution, we define $\vec X\in\mathbb{R}^{2N_p+2}$ 
as the $N_p+1$ first real components (denoted by $\Re$) of $X(\omega_k)$ 
followed by their imaginary components ($\Im$):
\begin{align}
    \vec X=&\left[\Re\left(X(\omega_0)\right),
      \dots,\Re\left(X(\omega_{N_p})\right),\right.\\
    &\qquad\left.\Im\left(X(\omega_0)\right), \dots, 
      \Im\left( X(\omega_{N_p})\right)\right].\nonumber
\end{align}
Note that the components $X_0$ and $X_{N_p+1}$ are the real
and imaginary DC components respectively (and that $X_{N_p+1}$ is
always zero).  Our mathematical system can thus be defined by
the nonlinear function $F: \mathbb{R}^{2N_p+3} \to
\mathbb{R}^{2N_p+2}$:
\begin{equation}
  \vec X = \vec F(\vec X,f_p).
\label{e:gensyst}
\end{equation}

Until now, the playing frequency has silently been assumed to be a known
quantity.  In autonomous systems, however, the frequency is an
additional unknown, so that the $N_p$-harmonic solution seeked is defined by
$2N_p{+}3$ unknowns linked through the $2N_p{+}2$ equations~(\ref{e:gensyst}).
However, it is well known that as
$\vec X$ is a periodic solution of a
dynamical system, any $\vec X^{\prime}$ deduced from
$\vec X$ by a phase rotation (i.e.\ a shift in
the time domain) is also a solution. Thus an additional constraint
has to be added in order to select a single periodic solution among
the infinity of phase-rotated solutions. A common choice (see Ref.\
\onlinecite{gilbert89}) is to consider the solution for which the 
first harmonic is real (i.e.\ its imaginary part, $X_{N_p+2}$, is zero). This
additional constraint decreases the number
of unknowns to $2N_p{+}2$ for an $N_p$-harmonic periodic
solution. Thus we get $\vec F: \mathbb{R}^{2N+2} \to \mathbb{R}^{2N+2}$, 
and it is now possible to find periodic solutions, if they
exist.

Finally, a simple way of avoiding trivial solutions to
equation~(\ref{e:gensyst}) is to look for roots of the function $\vec G: 
\mathbb{R}^{2N_p+2} \to  \mathbb{R}^{2N_p+2}$, defined by  
\begin{equation}
\vec G(\vec X,f_p)=\frac{\vec X-\vec F(\vec X,f_p)}{X_1},
\label{e:G}
\end{equation}
i.e.\ $\vec G(\vec X,f_p)=0$.  This equation is usually solved numerically
through an iteration process, for instance by the Newton-Raphson
method as in our case.  How to handle the playing frequency $f_p$ will
be discussed in the following section.

\subsection{Iteration by Newton-Raphson}\label{s:newton}

The equation $\vec G(\vec X,f) = 0$, $\vec G$ being defined by
equation~\eqref{e:G}, is nonlinear and has usually no analytical
solution.  (For readability we leave out the index $p$ on the
  playing frequency until end of Sec.~\ref{s:harmbal}.)  This
section describes the common, iterative Newton-Raphson method.
This is the method used in the program Harmbal (see
Section~\ref{s:harmbal}) although it had to be refined with a
backtracking procedure to improve its convergence, as discussed in
Section~\ref{s:backtracking}.

\begin{figure}
\ifgalleyfig   
  \includegraphics[width=1.95in]{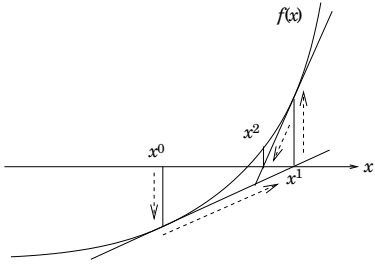}
\else
  \ifoutputfig  
  \includegraphics[width=5.5in]{\psdir/newton_g.eps}
  \fi
\fi
\caption
[The iteration process of Newton's method]
{\label{f:newton}
\ifgalleyfig
{The iteration process of Newton's method}
\fi}
\end{figure}

For the sake of later reference, it is useful to re\-col\-lect the
principles of Newton's method for a one-dimensional problem $g(x)=0$.
Starting with an estimate $x^0$ of the solution, the 
next estimate $x^1$ is defined as the intersection
point between the tangent to $g$ at $x_0$ and the 
$x$-axis. The method can be summarized as
\begin{equation}
        x^{i+1}=x^i-\frac{g(x^i)}{g'(x^i)}.
\end{equation}
This is repeated, as shown in Figure~\ref{f:newton}, while increasing
the iteration index $i$ until $g(x^i)<\varepsilon$, where
$\varepsilon$ is a user-defined threshold value.

In our $2N_p{+}2$-dimensional case, we have a vector
problem: we search $(\vec X,f)$ for which $\vec G(\vec X,f)=0$.  Newton's
method is generalized to the Newton-Raphson method,
which may be written \cite{numrec}:
\begin{equation}
        (\vec X^{i+1}, f^{i+1})=(\vec X^i , f^i )
        -\left(\mathbf{J}_G^i\right)^{-1}\!\!\cdot \vec G(\vec X^i,f^i),
\label{e:fullstep}
\end{equation}
where $\mathbf{J}^i_G{\triangleq}\nabla G(\vec X^i,f^i)$ is the
{\em Jacobian\/} matrix of $\vec G$ at $(\vec X^i,f^i)$. Note that 
all derivatives by $X_{N_p+2}$, which was chosen to be zero, are
ignored. The column $N_p{+}2$ in the Jacobian is thus replaced by the
derivatives with respect to the playing frequency $f$.
$\mathbf{J}^i_G$ is thus a $(2N_p{+}2)$-square
matrix. This means that line number $N_p{+}2$ in
equation~\eqref{e:fullstep} gives the new frequency $f$ instead of
$X_{N_p+2}$.
We define the {\em Newton step\/} $\Delta\vec X{=}\vec X^{i+1}{-}\vec
X^i$ (where $\Delta f{=}f^{i+1}{-}f^i$ replaces $\Delta X_{N_p+2}$),
which follows the local steepest descent direction.


The Jacobian may be found analytically if $\vec G$ is given analytically,
but it is usually sufficient to use the first-order approximation
\begin{equation}
  J_{jk}=\frac{\partial G_j}{\partial X_k}
        \simeq\frac{G_j(\vec X+\delta\vec X_k,f)-G_j(\vec X,f)}{\delta X},
\label{e:jacobij}
\end{equation}
except for $k=N_p+2$, in which case we use
\begin{equation}
  J_{j,N+2}=\frac{\partial G_j}{\partial f}
        \simeq\frac{G_j(\vec X,f+\delta f)-G_j(\vec X,f)}{\delta f}.
\label{e:jacobim}
\end{equation}
The components of $\delta\vec X_k$ are zero except for the $k$th
one, which is the tiny perturbation $\delta X$.  The iteration has
converged when $|\vec G^i|{\triangleq}|\vec G(\vec
X^i,f^i)|<\varepsilon$. 
We found $\varepsilon=10^{-5}$ to be a good compromize between
computation time and solution accuracy.

\section{Implementation and Harmbal}\label{s:harmbal}

\subsection{Equations for self-sustained musical
  instruments \label{s:self-sustained} }

Though, to the authors' knowledge, the harmonic balance method in the
context of musical acoustics with unknown playing frequency has only been applied to study models of
clarinet-like instruments, it should be possible to consider many
different classes of self-sustained instruments. 
It is well accepted that sound production by a musical instrument
results from the interaction between an exciter and a resonator
through a nonlinear coupling. Moreover, in most playing conditions,
linear modelling of both the exciter and the resonator is a good
approximation.

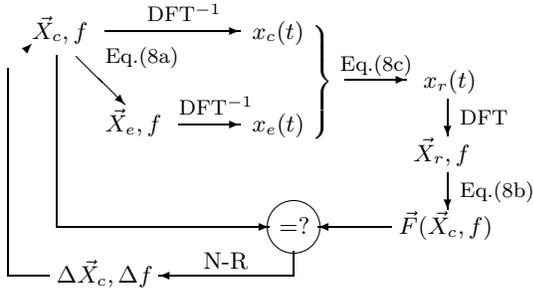
\begin{figure}
\centerline%
{
{\setlength{\unitlength}{1em}%
  \def\Xr{$\vec X_r,f$}\def\Xc{$\vec X_c,f$}\def\Xe{$\vec X_e,f$}
  \def\xr{$x_r(t)$}\def\xc{$x_c(t)$}\def\xe{$x_e(t)$}
\begin{picture}(22.5,12)(-1.5,-0.5)
\put(0,10){\Xc}
  \put(1.8,9.3){\vector(1,-1){2}}
  \put(3,9){\footnotesize Eq.(\ref{e:any_instr}a)}
  \put(3,6.2){\Xe}
    \put(6,6.5){\vector(1,0){2.5}}
    \put(6,6.9){\footnotesize DFT$^{-1}$}
    \put(9,6.2){\xe}
  \put(3,10.3){\vector(1,0){5.5}}
  \put(4.7,10.7){\footnotesize DFT$^{-1}$}
  \put(9,10){\xc}
\put(11.3,8){$\left.\hbox{\vbox to2.8\unitlength{}}\right\}$}
  \put(12.8,8.3){\vector(1,0){2.5}}
  \put(12.6,8.7){\footnotesize Eq.(\ref{e:any_instr}c)}
  \put(16,8){\xr}
  \put(17,7.5){\vector(0,-1){1.5}}
  \put(17.5,6.5){\footnotesize DFT}
  \put(15.6,5){\Xr}
  \put(17,4.5){\vector(0,-1){1.5}}
  \put(17.5,3.5){\footnotesize Eq.(\ref{e:any_instr}b)}
  \put(15,2){$\vec F$(\Xc)}
  \put(14.7,2.3){\vector(-1,0){3}}
\put(1,9.3){\line(0,-1){7}}
  \put(1,2.3){\vector(1,0){8.7}}
\put(10.7,2.3){\circle{2}}
\put(10.1,2){=?}
\put(10.7,1.3){\line(0,-1){1}}
\put(10.7,0.3){\vector(-1,0){5.5}}
\put(1,0){$\Delta\vec X_c,\Delta f$}
\put(7,.6){\small N-R}
\put(0.7,0.3){\line(-1,0){1.7}}
\put(-1,0.3){\line(0,1){8.5}}
\put(-1,8.8){\vector(1,1){1}}
\end{picture}}}
\caption{The iteration loop of the harmonic balance method for a
musical instrument (notations defined in the text)}
\label{f:any_instr_solve}
\end{figure}

Therefore, within these hypotheses, any musical instrument could be
modelled by the following three equations:
\begin{equation}
\qquad\quad\left\{
\begin{array}{l@{\qquad}c}
Z_e(\omega) X_e(\omega) = X_c(\omega)&    \mathrm{(a)}\\
\phantom{Z_e(\omega)}X_c(\omega) = Z_r(\omega) X_r(\omega) &   \mathrm{(b)}\\
\mathcal{F}(x_c(t),x_e(t),x_r(t)) =0 & \mathrm{(c)}
\end{array}
\right.
\label{e:any_instr}
\end{equation}
where $Z_e$ is the dynamic stiffness and $Z_r$ is the input impedance of the exciter and the
resonator, respectively, and $X_e$ and $X_r$ are the spectra
describing the dynamics of the exciter and the resonator during the
steady state (periodicity assumption).  $X_c$ is the spectrum of
the coupling variable.  All these quantities, and thus
equations~(\ref{e:any_instr}a--b), are defined in the
Fourier domain. Equation~(\ref{e:any_instr}c) is
written in the time domain, where $\mathcal{F}$ is a nonlinear functional of
$x_c$, $x_e$, and $x_r$, which are the inverse Fourier transforms of
$X_c$, $X_e$, and $X_r$, respectively.  We apply the discretization as
described in Section~\ref{s:HBM}, implying that
equations~(\ref{e:any_instr}a--b) become vector equations where the
impedances must be written as real $(2N_p{+}2){\times}(2N_p{+}2)$-matrices to accommodate the rules of complex multiplication:
\begin{equation}
  Z(f)=\left(
  \begin{array}{cc}
    \Re(\althat{Z}(f))&-\Im(\althat{Z}(f))\\
    \Im(\althat{Z}(f))&\Re(\althat{Z}(f))\\
  \end{array}
  \right)
\label{e:impmat1}
\end{equation}
where
\begin{equation}
  \althat{Z}(f)=\left(
  \begin{array}{cccc}
    Z(0) &0&\cdots&0\\
    0& Z(\omega_1)&&0\\
    \vdots&&\ddots&\vdots\\
    0& 0 &\cdots&Z(\omega_{N_p})
  \end{array}
  \right)
\label{e:impmat2}
\end{equation}
is complex, and $\Re(\althat{Z})$ and $\Im(\althat{Z})$ are the
real and imaginary components of $\althat{Z}$.
The system~(\ref{e:any_instr}) is solved iteratively
by Harmbal according to the scheme illustrated in Figure~\ref{f:any_instr_solve}.

In Harmbal, these equations are easily defined by writing new C
functions. Only superficial knowledge of the C language is necessary
to do this.

Three cases related to models of single reed instruments with
cylindrical or stepped-conical bores are studied in particular in
Section~\ref{s:case} in order to validate the code and to illustrate the
modularity of Harmbal.

\subsection{Practical characteristics of Harmbal \label{s:harmdet}}
Both fast calculation, good portability, and independence of
commercial software are easily achieved by programming in C, whose
compiler is freely available for most computer platforms.  It is,
however, somewhat difficult to combine portability with easy usage,
because an intuitive usage normally means a graphical and interactive
user interface, while the handling of graphics varies a lot between the
different platforms.

We have chosen to write Harmbal with a nongraphical and
non-interactive\footnote{The term {\em non-interactive\/}
means that the user has no influence on the program while it is
running.} user interface.  The major advantage of this is that
independent user interfaces may be further developed depending on need.

Our concept is to save both the parameters and the solution in a
single file.  This file also serves as input to Harmbal while individual
parameters can be changed through start-up arguments.  The solution
provided by the file works as the initial condition for the harmonic
balance method.  Thus the lack of a simple user interface is
compensated by a simple way of re-using an existing solution to solve
the system for a slightly different set of parameters.  Solutions for
a range of a parameter values may thereby be calculated by changing the
parameter stepwise and providing the previous solution as an
initial
condition for the next run. The Perl script {\em hbmap\/} provides such zeroth-order continuation facilities.
This procedure may
also be used when searching for a solution where it is difficult to
provide a sufficiently good initial condition, for instance by
successively increasing $N_p$ when wanting many harmonics.

\subsection{Convergence of Newton-Raphson \label{s:holes}}

\begin{figure}
\ifgalleyfig%
  \includegraphics[width=3.25in]{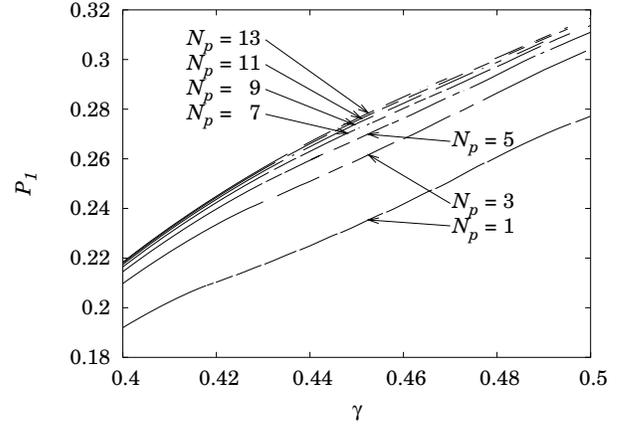}%
\else
  \ifoutputfig%
  \includegraphics[width=5.5in]{\psdir/compmap.eps}%
  \fi
\fi
\caption
[Solution holes: first pressure harmonic
$P_1$ versus blowing pressure $\gamma$ for different $N_p$ with
$N_t=128$, $\zeta=0.5$, and $\eta=10^{-3}$.  (Even $N_p$ give the same as
$N_p{-}1$.) Equations and parameters are defined in section \ref{s:case}.]
{\label{f:holes}
\ifgalleyfig
{Solution holes: first pressure harmonic
$P_1$ versus blowing pressure $\gamma$ for different $N_p$ with
$N_t=128$, $\zeta=0.5$, and $\eta=10^{-3}$.  (Even $N_p$ give the same as
$N_p{-}1$.) Equations and parameters are defined in section \ref{s:case}.}
\fi}
\end{figure}

When merely employing the Newton-Raphson method to determine the
solution of the system at a given set of parameters, we have found
that it is impossible to find a solution at particular combinations of
the parameters.  Indeed, for the clarinet model of
Section~\ref{s:helcyl}, no convergence was obtained for particular
values of the parameter $\gamma$ (the dimensionless blowing pressure)
and its neighborhood. This is seen as discontinuities, or
\emph{holes}, in the curves in Figure~\ref{f:holes} 
(see Section~\ref{s:case} for the underlying equations and
parameters). Note that the  
solutions seem to go continuously through this hole and that the
positions of the holes and their extent vary with the number of
harmonics $N_p$ taken into account.
The curves were calculated by the program {\em hbmap}. In this case
we have decreased $\gamma$ from 0.5 downward in steps of $10^{-4}$ and
drawn a line between them except across $\gamma$ values where solution failed.
In the holes, the Newton-Raphson 
method did not converge, either by alternating between two values of
$\vec P$ (i.e.\ $\vec X_c$)  or by starting to diverge.

\begin{figure}
\ifgalleyfig%
  \includegraphics[width=3.25in]{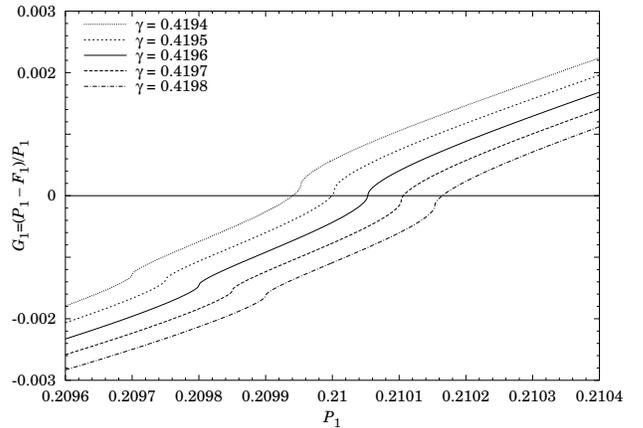}%
\else
  \ifoutputfig%
  \includegraphics[width=5.5in]{\psdir/jumphole128.eps}%
  \fi
\fi
\caption
[$G_1$ as $P_1$ varies around the solution $G_1=0$ for various
$\gamma$ around a hole at $\gamma\simeq0.4196$.  $N_t=128$ and
  $N_p=1$.]
{\label{f:GvsP}
\ifgalleyfig
{$G_1$ as $P_1$ varies around the solution $G_1=0$ for various
$\gamma$ around a hole at $\gamma\simeq0.4196$.  $N_t=128$ and
  $N_p=1$.}
\fi}
\end{figure}

To study the problem, we simplified the system to a one-dimensional
problem by setting $N_p=1$, thus leaving $P_1$ as
the only nonzero value.  $G_1$ thus became the only contributor to
$|\vec G|$, and a simple graph of $G_1$ around the
solution $G_1=0$ could illustrate the problem, as shown in
Figure~\ref{f:GvsP}.
We see that the curve of $G_1(P_1)$ has inflection points
(visible as ``soft steps'' on the curve) at rather regular
distances. At the centre of a convergence hole, 
i.e.\ for $\gamma\simeq0.4196$, an inflection point is located at the
intersection with the horizontal axis.  This is a school example of
a situation where Newton's method does not converge because the Newton
step $\Delta P_1$ brings us alternatingly from one side of the
solution to the other, but not closer.

\begin{figure*}
\ifgalleyfig%
  \includegraphics[height=.595\width]{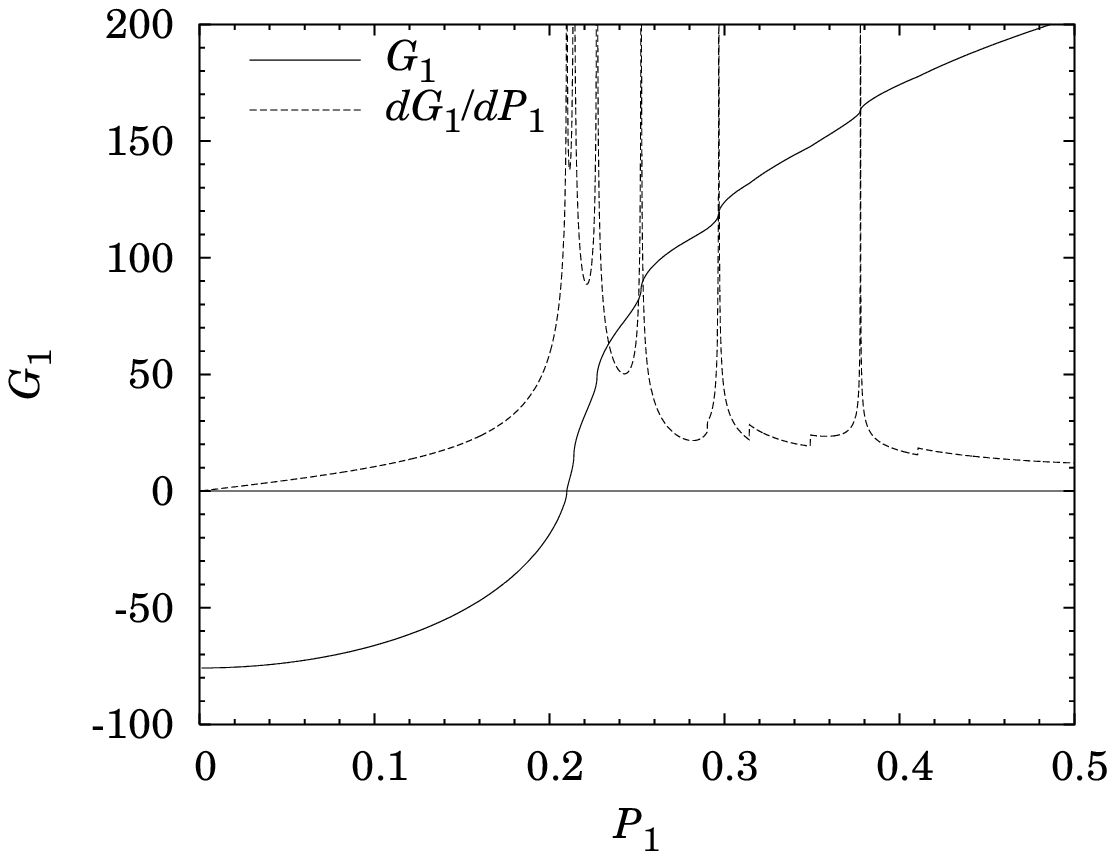}\hspace{-.21\width}%
  \includegraphics[height=.595\width]{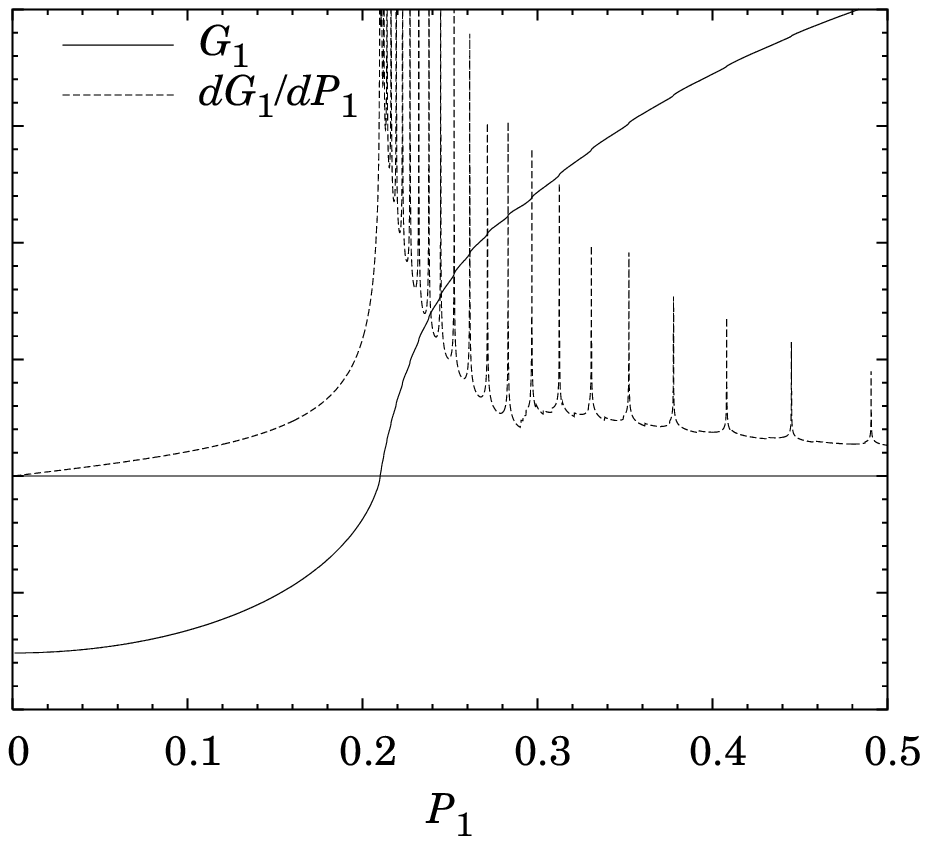}\hspace{-.21\width}%
  \includegraphics[height=.595\width]{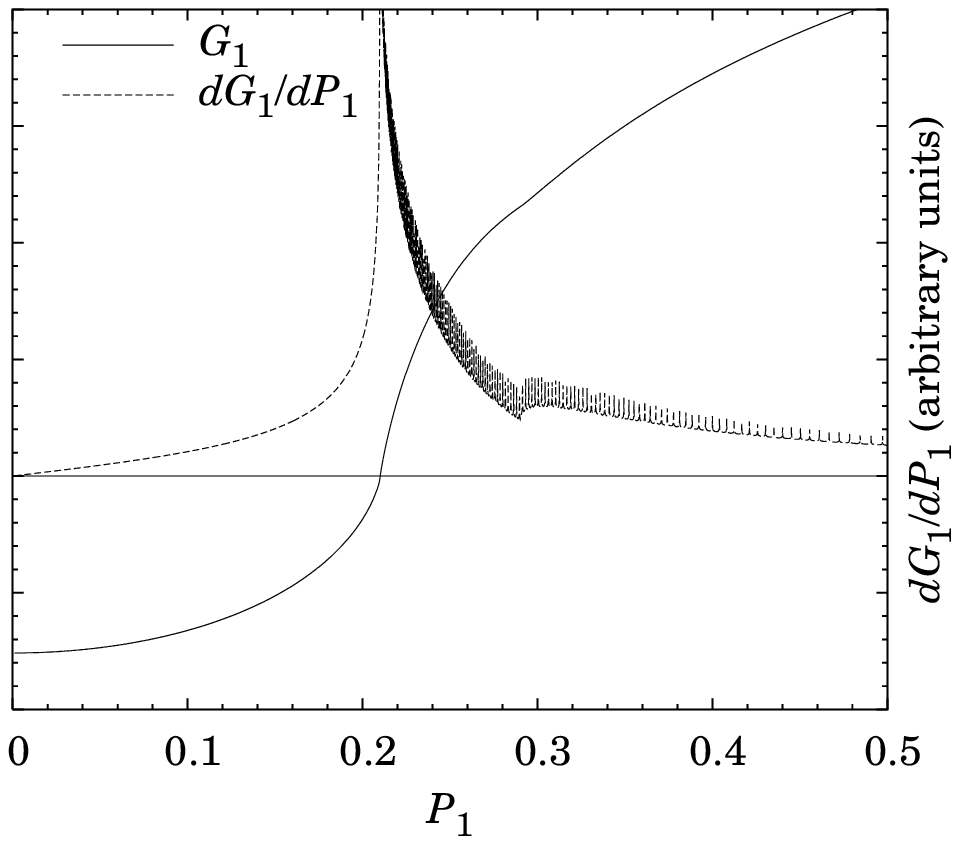}\\
\else
  \ifoutputfig%
  \includegraphics[height=.285\width]{\psdir/jumps32.eps}\hspace{-.1\width}%
  \includegraphics[height=.285\width]{\psdir/jumps128.eps}\hspace{-.1\width}%
  \includegraphics[height=.285\width]{\psdir/jumps1024.eps}%
  \fi
\fi
\caption
[The effect of sampling rate on the ``smoothness'' of $G_1(P_1)$:
(a) $N_t=32$, (b) 128, and (c) 1024.  The derivative $dG_1/dP_1$
exhibits the ``roughness''.]
{\label{f:sampling}
\ifgalleyfig
{The effect of sampling rate on the ``smoothness'' of $G_1(P_1)$:
(a) $N_t=32$, (b) 128, and (c) 1024.  The derivative $dG_1/dP_1$
exhibits the ``roughness''.}
\fi}
\end{figure*}

In fact, the existence of inflection points is linked with the digital
sampling of the continuous signal.  If the sampling rate is increased,
i.e.\ if $N_t$ is increased, the steps become smaller but occur more
frequently, as shown for $N_t=32$, 128, and 1024 in
Figures~\ref{f:sampling}a--c.  The derivative $dG_1/dP_1$ is included
in the figures to quantify the importance of the steps.  According to
the Figures~\ref{f:sampling}a--c it seems reasonable to increase $N_t$
to avoid convergence problems. However, this would significantly
increase the computational cost. Another solution is therefore
suggested in the following.

\subsection{Backtracking}\label{s:backtracking}

\begin{figure}
\ifgalleyfig%
  \includegraphics[width=2.4in]{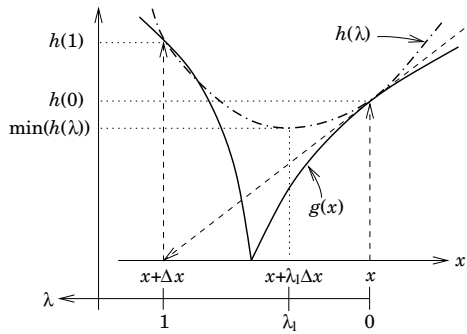}%
\else
  \ifoutputfig%
  \includegraphics[width=5.5in]{\psdir/backtrack3.eps}%
  \fi
\fi
\caption
[The principle of backtracking in one dimension. Objective is
  to estimate the root of $g(x)$ (solid curve). Broken lines with arrows
  show how the Newton step $\Delta x$ from $x$ leads to divergence.
  $h(\lambda)$ (dot-dashed curve) is a 2nd order expansion of $g(x)$
  along the Newton step, i.e. the $\lambda$ axis.  Minimum of
  $h(\lambda)$ should be closer to the root of $g(x)$ than $g(x+\Delta
  x)$.]
{\label{f:backtracking}
\ifgalleyfig
{The principle of backtracking in one dimension. Objective is
  to estimate the root of $g(x)$ (solid curve). Broken lines with arrows
  show how the Newton step $\Delta x$ from $x$ leads to divergence.
  $h(\lambda)$ (dot-dashed curve) is a 2nd order expansion of $g(x)$
  along the Newton step, i.e. the $\lambda$ axis.  Minimum of
  $h(\lambda)$ should be closer to the root of $g(x)$ than $g(x+\Delta
  x)$.}
\fi}
\end{figure}

When the Newton-Raphson scheme fails to converge, it often happens
because the Newton step $\Delta\vec X$ leads to a point where $|\vec G(\vec
X,f)|$ is larger than in the previous step.  However, acknowledging
that the Newton step points in the direction of the steepest descent, there
must be a point along $\Delta\vec X$ where $|\vec G(\vec X,f)|$ is smaller
than in the previous iteration of the HBM.  A backtracking
algorithm described in Numerical Recipes \cite[Sec.9.7]{numrec} solves the
problem elegantly by shortening the Newton step as 
described here.  The principle is illustrated in the simple
one-dimensional case in Figure~\ref{f:backtracking}, where $g(x)$
replaces $|\vec G(\vec X,f)|$, although we use the multidimensional
notation in the following.
Defining the $\lambda$ axis along the Newton step, we simply take a
step $\lambda\Delta\vec X$ in the 
same direction, where $0<\lambda<1$. The optimal value for $\lambda$
is the one that minimizes the function $h(\lambda)$:
\begin{equation}
        h(\lambda)=\textstyle\frac12|\vec G(\vec X^i+\lambda\Delta\vec X)|^2
\end{equation}
with derivative
\begin{equation}
        h'(\lambda)=\left(\mathbf{J}_G \cdot \vec G\right)\big|
                _{\vec X^i+\lambda\Delta\vec X} \cdot\Delta\vec X.
\end{equation}
During the calculation of the failing Newton step, we computed $\vec
G(\vec X^i)$ and $\vec G(\vec X^{i+1})$, so now it is possible to
calculate with nearly no additional computational effort $h(0) =
\frac12|\vec G(\vec X^i)|^2$, $h'(0) = -|\vec G(\vec X^i)|^2$, and
$h(1) = \frac12|\vec G(\vec X^i+\Delta\vec X)|^2 = \frac12|\vec G(\vec
X^{i+1})|^2$. This allows to propose a quadratic approximation of $h$
for $\lambda$ between $0$ and $1$, for which the minimum is located at
\begin{equation}
        \lambda_1=-\frac{\frac12h'(0)}{h(1)-h(0)-h'(0)}.
\end{equation}
It can be shown that $\lambda_1$ should not exceed 0.5, and in practice
$\lambda_1\ge0.1$ is required to avoid a too short step at
this stage.

If $|\vec G(\vec X^i+\lambda_1\Delta\vec X)|$ still is larger than
$|\vec G(\vec X^i)|$, $h(\lambda)$ is then modelled as a cubic function (using
$h(\lambda_1)$ which has just been calculated). The minimum of this
cubic function gives a new value $\lambda_2$, again restricted to
$0.1\lambda_1<\lambda_2<0.5\lambda_1$.  This 
calculation requires solving a system of two equations, so if also
$\lambda_2$ is not accepted because 
$|\vec G(\vec X^i+\lambda_2\Delta\vec X)|$ is still too large, we do
not enhance to a fourth-order model of $h$, which would increase the
computational cost much more. Instead, subsequent cubic modellings are
performed using the most two recent values of $\lambda$.  In practice,
however, not many repetitions should be necessary before finding a
better solution, if possible.

\section{Case studies}\label{s:case}

\subsection{The equations for the clarinet \label{s:clarinet}}


The three equations~(\ref{e:any_instr}a--c) may be constructed by
physical modelling. In the case of the clarinet, a common simple model
is described below.  We limit the description in the following to a
brief presentation based on dimensionless quantities, {\em
  dimensional\/} variables being denoted by a hat ($\dimvar{\ }$)
hereafter (see Fritz et al.\ \cite{fritz04} for further details).

\begin{figure}
\ifgalleyfig%
  \includegraphics[width=1.9in]{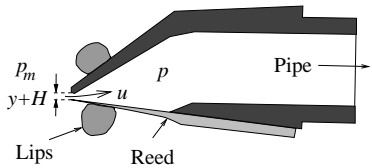}%
\else
  \ifoutputfig%
  \includegraphics[width=5.5in]{\psdir/mouthpiece.eps}
  \fi
\fi
\caption
[Illustration of the mouthpiece]
{\label{f:mouthpiece}
\ifgalleyfig
{Illustration of the mouthpiece}
\fi}
\end{figure}

\medskip
The exciter is an oscillating reed which may be modelled as a spring
with mass and damping:
\begin{equation}
        \ddot{\dimvar{y}}+g_e\dot{\dimvar{y}}+\omega_e^2\dimvar{y}
          =\frac1{\mu_e}(\dimvar{p}-p_m),
\label{e:lindiff_dim}
\end{equation}
where $\dimvar{y}$ is the dynamic reed displacement, and $\dimvar{p}$
and $p_m$ are the dynamic pressure in the mouthpiece, i.e.\ the {\em
  internal\/} pressure, and the static blowing pressure in the player's
mouth, respectively.  The constants $\mu_e$, $g_e$, and
$\omega_e$ represent the mass per area, the damping factor, and the
angular resonance frequency of the exciter (the reed).  The dots over
$\dimvar{y}$ denote the time derivative.  In dimensionless form,
equation~\eqref{e:lindiff_dim} becomes
\begin{equation}
        M\ddot{x}+R\dot{x}+Kx=p,
\label{e:lindiff}
\end{equation}
where $p=\dimvar p/p_M$ and $x=\dimvar y/H+\gamma/K$ with
$\gamma=p_m/p_M$.  The equilibrium reed opening is $H$ as shown in
Figure~\ref{f:mouthpiece}. In the static regime, when
blowing harder than a maximum pressure $p_M$, i.e.\
$p_m\ge p_M$ ($\gamma\ge1$),  the reed blocks the opening, i.e.\
$\dimvar y=-H$, so we get $\dimvar p=0$ and can conclude that $K=1$ for
the current reed model.

Like Fritz et al.\ \cite{fritz04} we relate the dimensionless time to
the resonance angular frequency $\omega_r$ of the resonator (the
bore), i.e.\ $t=\omega_r\dimvar t$, so that the values of the
dimensionless mass $M$, damping $R$, and spring constant $K$ become
\begin{align}
  K&=\mu_e H\omega_e^2/p_M=1,\\
  R\,&=Kg_e\omega_r/\omega_e^2=g_e\omega_r/\omega_e^2,\\ 
  M&=K\omega_r^2/\omega_e^2=\omega_r^2/\omega_e^2.
\label{e:MRK}
\end{align}
In the Fourier domain, Equation~\eqref{e:lindiff} thus takes the form of
equation~(\ref{e:any_instr}a), $Z_e(\omega)X(\omega)=P(\omega)$, where
\begin{equation}
  Z_e(\omega)=1-M\left(\!\frac\omega{2\pi}\!\right)^{\!2}
      +iR\left(\!\frac\omega{2\pi}\!\right),
\label{e:reedimp}
\end{equation}
for $i=\sqrt{-1}$ and $\omega =2\pi\dimvar\omega/\omega_r
=\dimvar\omega/f_r$ is the dimensionless angular frequency in the Fourier
domain.

A common minimum model for the clarinet assumes a simple reed with no
mass or damping, thus $M=R=0$. Equation~\eqref{e:lindiff} reduces to
$x=p$. 

\medskip 
The resonator (i.e.\ the air column in the bore of the
instrument) is commonly described by its frequency response $\dimvar
Z_r(\dimvar\omega)$. For a simple cylindrical bore of length $l$ with
a closed and an ideal open end, the resonance frequencies are odd
multiples of $f_r=c/4l$, $c$ being the sound speed in the air column
\cite{fletcher91}.  The input impedance of the bore may thus be
expressed in dimensionless quantities as
\begin{equation}
  Z_r(\omega)=\frac{\dimvar Z_r(\dimvar\omega)}{Z_0}
  =i\tan\!\left(\frac\omega4 + (1-i)\alpha(\omega)\!\right),
\label{e:freqresponse}
\end{equation}
where $\alpha(\omega)\triangleq\psi\eta\sqrt{\omega/2\pi}$ with
$\psi\simeq1.3$ for common conditions in air and $\eta$ being the
dimensionless loss parameter, which depends on the tube length,
typically 0.02 for a normal clarinet with all holes closed.
$Z_0\triangleq \rho c/S$ is the characteristic impedance of the
cylidrical resonator, $S$ being its cross section, and $\rho$ the
density of air.  The last term in the argument of
equation~\eqref{e:freqresponse} includes the dispersion as the real
part and viscous losses as the imaginary part.

Equation~(\ref{e:any_instr}b) becomes
\begin{equation}
        P(\omega)=Z_r(\omega)U(\omega),
\label{e:imped}
\end{equation}
where $P(\omega)$ and $U(\omega)\triangleq\dimvar U(\omega)Z_0/p_M$ are
the dimensionless internal pressure and volume flow of air through the
mouthpiece in the Fourier domain.

\medskip
The coupling equation~(\ref{e:any_instr}c), is given by the Bernoulli
theorem with some supplementary hypotheses applied
between the mouth and the outlet of the reed channel. 
 The coupling equation is nonlinear and must be calculated
in the time domain. This leads to the following expression for the
dimensionless airflow  through the mouthpiece
\cite{kergomard95}:
\begin{equation}
        u(p,x)=\zeta\left(1+x-\gamma\right)
                \sqrt{|\gamma-p|}\sign(\gamma-p)
\label{e:nonlin}
\end{equation}
as long as $x>\gamma-1$, and $u=0$ otherwise.
$\zeta=Z_0wH\sqrt{2/\rho p_M}$ is a dimensionless embouchure parameter
roughly describing the mouthpiece and the position of the player's
mouth, $w$ being the width of the opening and $\rho$ the density of
the air. $\zeta$ is also related to the maximum volume velocity
entering the tube \cite{ollivier04a}.

If the reed dynamics were not taken into account, we had $x=p$ and thus
\begin{equation}
        u(p)=\zeta\left(1+p-\gamma\right)
                \sqrt{|\gamma-p|}\sign(\gamma-p)
\label{e:simplenonlin}
\end{equation}
for $p>\gamma-1$, and, as before, $u=0$ otherwise.


\subsection{Verification of method and models \label{s:verif}}

In the following we want to verify that the HBM (and its
implementation in Harmbal) gives correct results.  By using very low
losses in the resonator (small $\eta$) we can compare the results of
the HBM with analytical results. Rising the attenuation in the resonator and
including mass and damping for the exciter, we compare with numerical
results from real-time synthesis of the same system.  This also gives us
the opportunity to illustrate the modularity of Harmbal as we change the models
of the resonator and the nonlinear coupling.

\subsubsection{Helmholtz oscillation for cylindrical tubes}\label{s:helcyl}

\begin{figure}[t]
\ifgalleyfig%
  \includegraphics[width=3.25in]{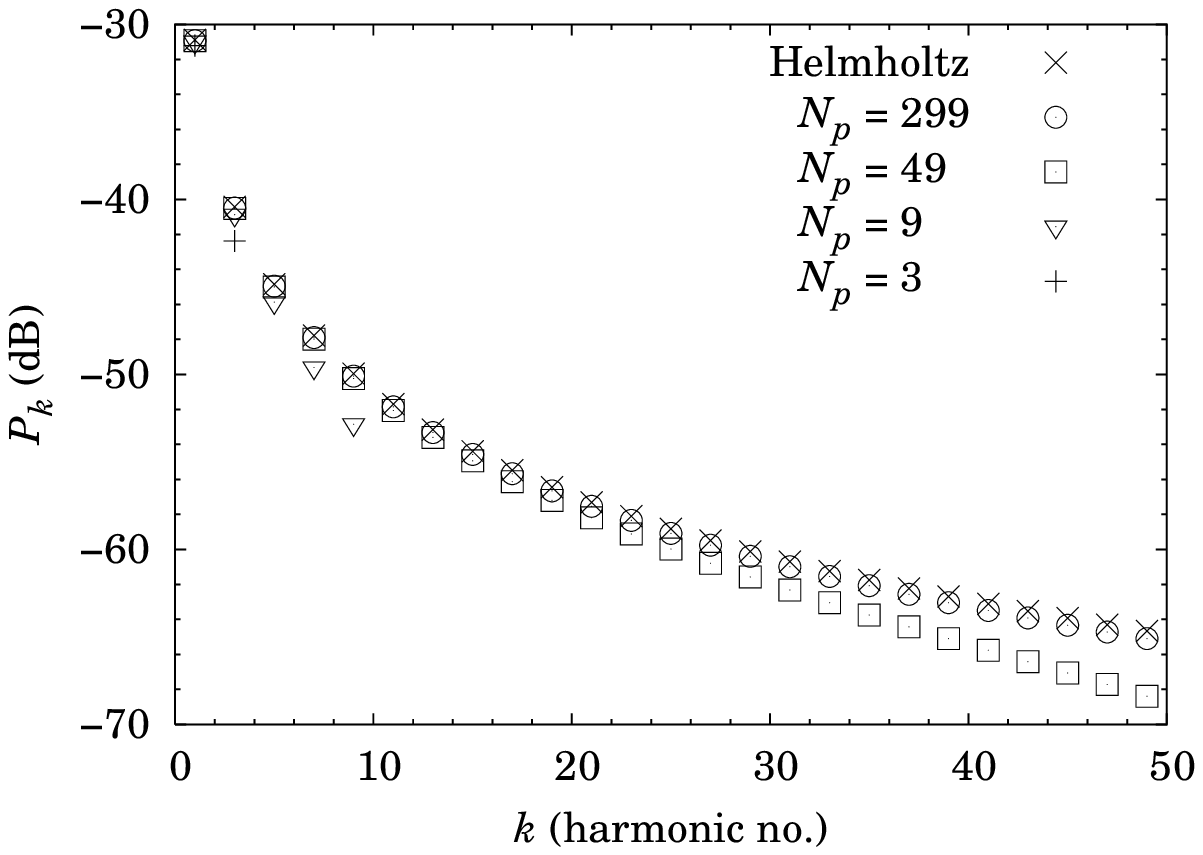}%
  \\\includegraphics[width=3.25in]{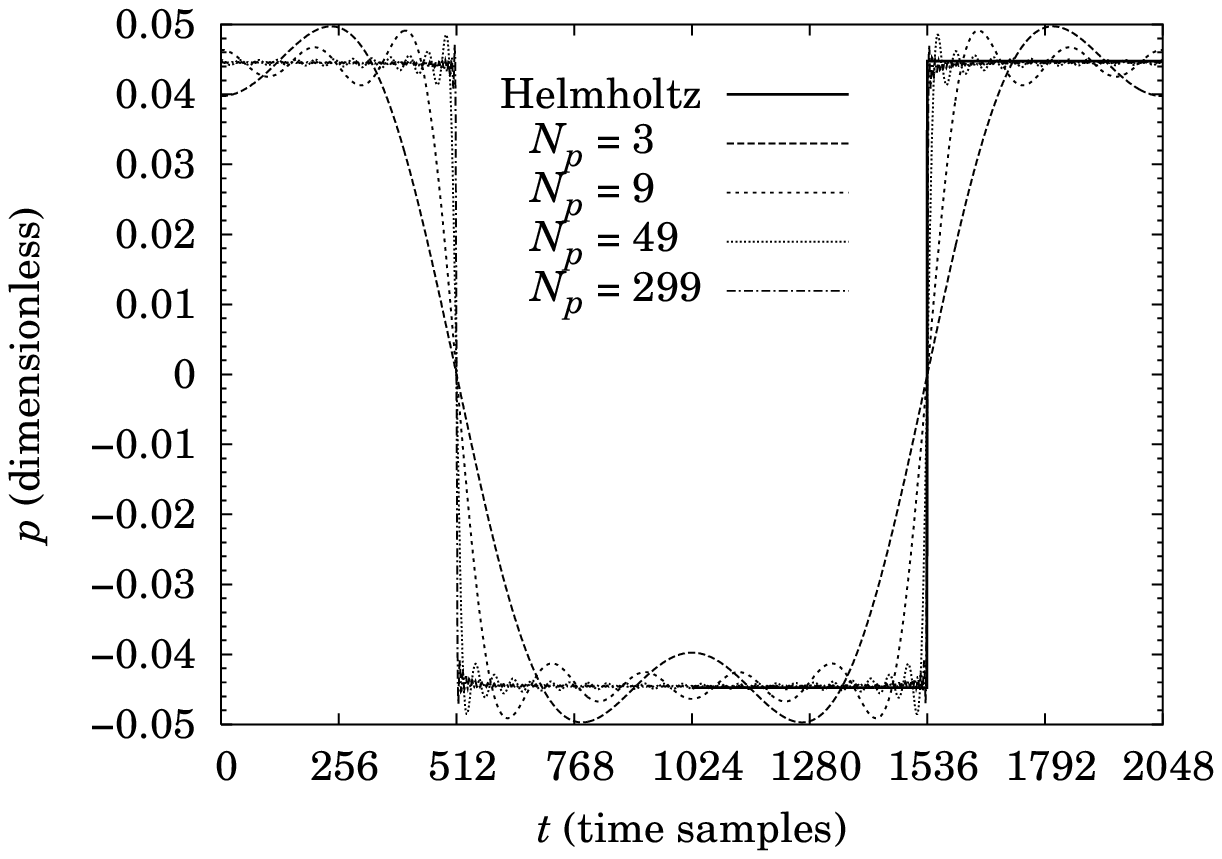}%
\else
  \ifoutputfig%
  \includegraphics[width=5.5in]{\psdir/HBM-hh-freq-g0.334336.eps}%
  \\\includegraphics[width=5.5in]{\psdir/HBM-hh-time-g0.334336.eps}%
  \fi
\fi
\caption
[The Helmholtz solution, eq.\
  \eqref{e:helmotion} compared with the HBM truncated to 3, 9, 49,
  and 299 harmonics close to the oscillation threshold
  ($\gamma=0.334$, $\zeta=0.5$, $\eta=10^{-5}$).  (a) frequency
  domain.  (b) time domain.]
{\label{f:nearthres-nl}
\ifgalleyfig
{The Helmholtz solution, eq.\
  \eqref{e:helmotion} compared with the HBM truncated to 3, 9, 49,
  and 299 harmonics close to the oscillation threshold
  ($\gamma=0.334$, $\zeta=0.5$, $\eta=10^{-5}$).  (a) frequency
  domain.  (b) time domain.}
\fi}
\end{figure}

\begin{figure}
\ifgalleyfig%
  \includegraphics[width=3.25in]{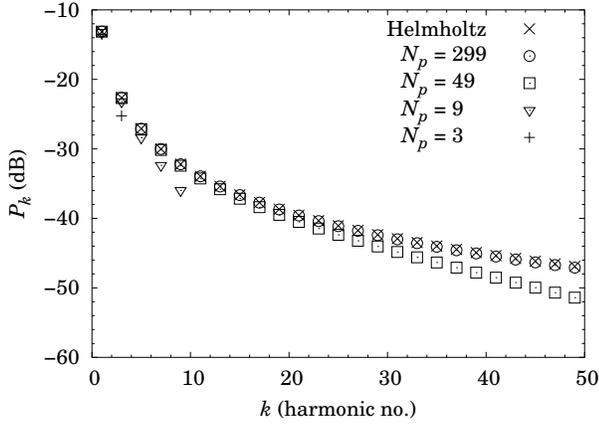}%
\else
  \ifoutputfig%
  \includegraphics[width=5.5in]{\psdir/HBM-hh-freq-g0.40.eps}%
  \fi
\fi
\caption
[The Helmholtz solution, eq.\
  \eqref{e:helmotion} compared with the HBM truncated to 3, 9, 49,
  and 299 harmonics far from the oscillation threshold ($\gamma=0.40$,
  $\zeta=0.5$, $\eta=10^{-5}$) in the frequency domain.]
{\label{f:largeosc-nl}
\ifgalleyfig
{The Helmholtz solution, eq.\
  \eqref{e:helmotion} compared with the HBM truncated to 3, 9, 49,
  and 299 harmonics far from the oscillation threshold ($\gamma=0.40$,
  $\zeta=0.5$, $\eta=10^{-5}$) in the frequency domain.}
\fi}
\end{figure}

To compare the HBM results with analytical results, we assume a
nondissipative, nondispersive air column, i.e.\ setting $\eta=0$ and
thus $\alpha=0$ in equation~\eqref{e:freqresponse}.  Furthermore, we
assume that the reed has neither mass nor damping and thus use
equation~\eqref{e:simplenonlin}.  The resulting square-wave amplitude
(the Helmholtz motion \cite{helmholtz77}) may be found by solving
$u(p)=u(-p)$, which results from the fact that the internal pressure
$p(t)$ and  the power $p(t)u(t)$ averaged over a period are zero
according to the lossless hypothesis \cite{kergomard95}.  This leads
to the square oscillation with amplitude
\begin{equation}
   p(\gamma)=\sqrt{-3\gamma^2+4\gamma-1}.
\label{e:helmotion}
\end{equation}
This result is compared with the results calculated by Harmbal (for the
same set of equations, but $\eta=10^{-5}$ instead of $\eta=0$ to avoid
infinite impedance peaks) for 3, 9, 49, and 299 harmonics
close to the oscillation threshold in
Figure~\ref{f:nearthres-nl}, and at $\gamma=0.4$ in
Figure~\ref{f:largeosc-nl}, which is far from the
threshold.  

As expected, the solution using the HBM shows good convergence towards
the Helmholtz motion as the number of harmonics increases.  Note the
deviation for higher
harmonics close to the threshold, even for 299 harmonics.
Dissipation in the resonator ($\eta=10^{-5} \not= 0$)
causes higher harmonics to be damped more in this area of $\gamma$
than for higher blowing pressures (as explained e.g.\ in Ref.\
\onlinecite{kergomard00}). The deviation from a square-wave 
signal is thus more noticeable close to the threshold, and as the HBM
calculations imposed a nonzero dissipation, this is probably the
reason for the small deviation in Figure~\ref{f:nearthres-nl}.  The
deviation is not visible in the time domain.

\begin{figure}
\ifgalleyfig%
  \includegraphics[width=3.25in]{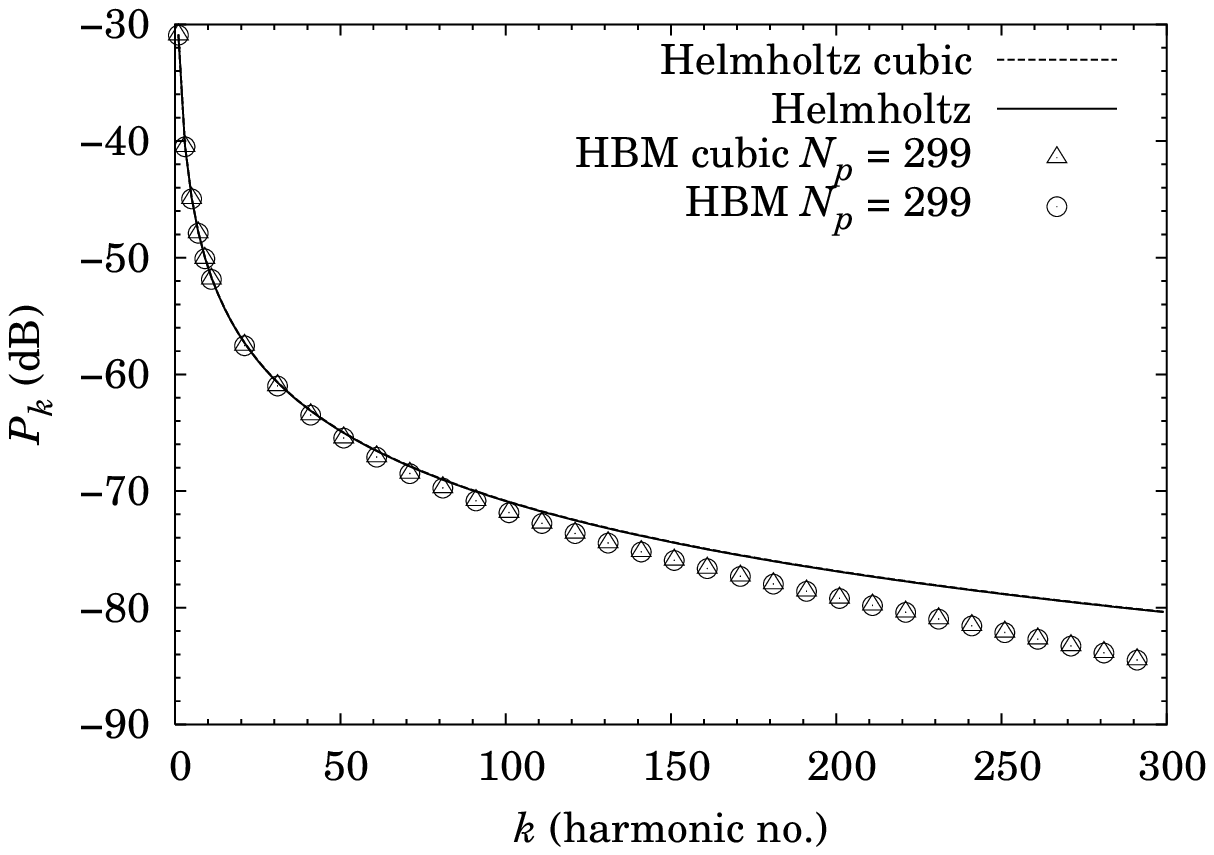}%
  \\\includegraphics[width=3.25in]{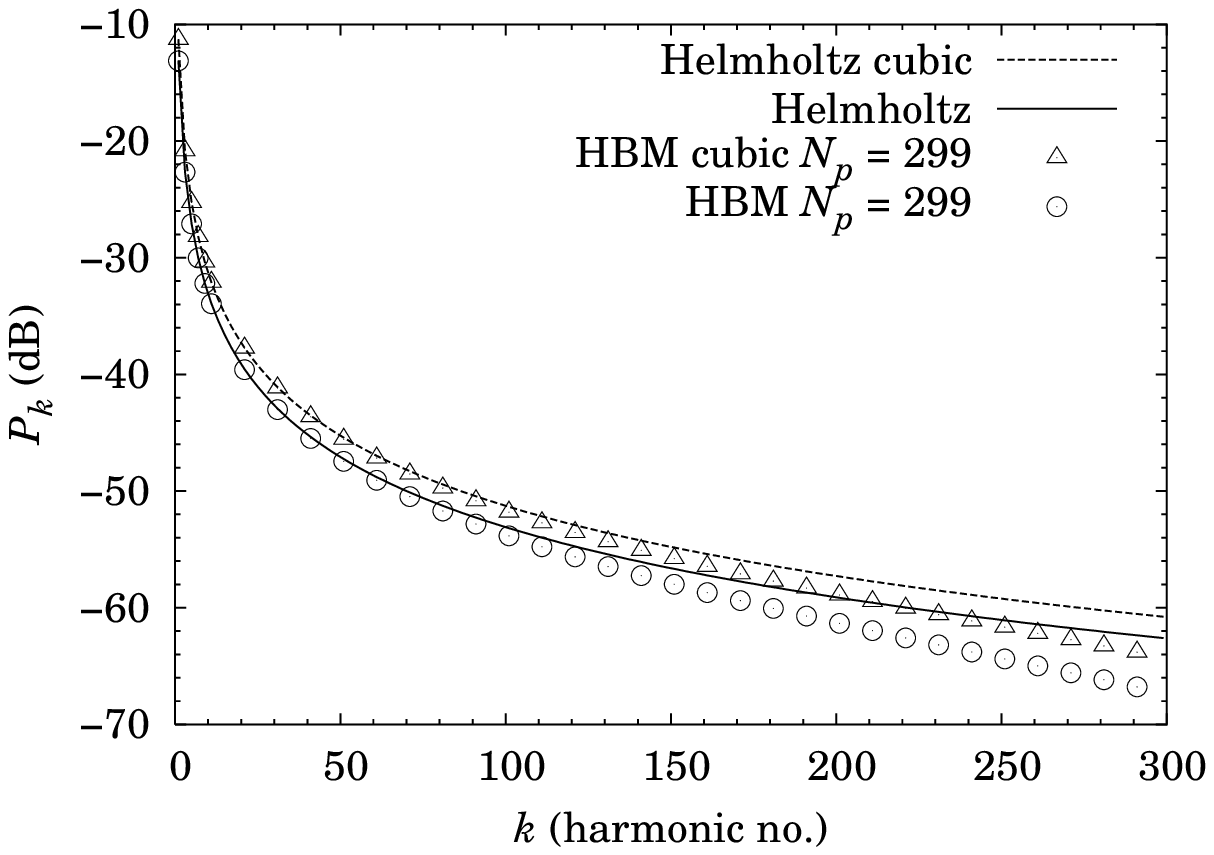}%
\else
  \ifoutputfig%
  \includegraphics[width=5.5in]{\psdir/compcub-freq-g0.334336.eps}%
  \\\includegraphics[width=5.5in]{\psdir/compcub-freq-g0.40.eps}%
  \fi
\fi
\caption
[The (lossless) Helmholtz motion and the
  (almost lossless) HBM for 299 harmonics using the full
  nonlinearity~\eqref{e:simplenonlin} and 
  the cubic expansion~\eqref{e:cubic} (a) close to the 
  oscillation threshold ($\gamma=0.334$) and (b) far from it
  ($\gamma=0.40$) for $\zeta=0.5$, $\eta=10^{-5}$. Above the 11th
  harmonic only every 10th harmonic is shown.]
{\label{f:compcub}
\ifgalleyfig
{The (lossless) Helmholtz motion and the
  (almost lossless) HBM for 299 harmonics using the full
  nonlinearity~\eqref{e:simplenonlin} and 
  the cubic expansion~\eqref{e:cubic} (a) close to the 
  oscillation threshold ($\gamma=0.334$) and (b) far from it
  ($\gamma=0.40$) for $\zeta=0.5$, $\eta=10^{-5}$. Above the 11th
  harmonic only every 10th harmonic is shown.}
\fi}
\end{figure}

A popular simplification of the nonlinear
function~\eqref{e:simplenonlin} is a cubic expansion for small
oscillations (e.g.\ Ref.\ \cite{mcintyre83,grand97,worman71}):
\begin{equation}
\althat{u}(p) = u_{00}+Ap+Bp^2+Cp^3,
\label{e:cubic}
\end{equation}
where $u_{00}$, $A$, $B$, and $C$ are easily found by expanding
equation~\eqref{e:simplenonlin}. Its Helmholtz solution is easily
calculated like above, yielding 
\begin{equation}
 p(\gamma)=\sqrt{-\frac AC}=\sqrt{\frac{8\gamma^2(3\gamma-1)}{\gamma+1}}.
\label{e:pcubic}
\end{equation}  
The influence of
the difference between the two versions of the nonlinear function is investigated in
Figure~\ref{f:compcub} for the lossless case.
Close to the oscillation threshold, Figure~\ref{f:compcub}a, we see
that there is no significant difference between the two versions of
the nonlinear equation, as expected.  The fact that the HBM is lower
for higher harmonics is as before due to the small attenuation we had to
include to perform the numerical calculations.  Far from the
threshold, however, Figure~\ref{f:compcub}b, we see that the cubic
expansion fails to approximate the nonlinear equation.  For lower
harmonics this error is larger than the attenuation effect in the HBM
calculations. This is further discussed by Fritz et al \cite{fritz04}.

\begin{figure}
\ifgalleyfig%
  \includegraphics[width=3.25in]{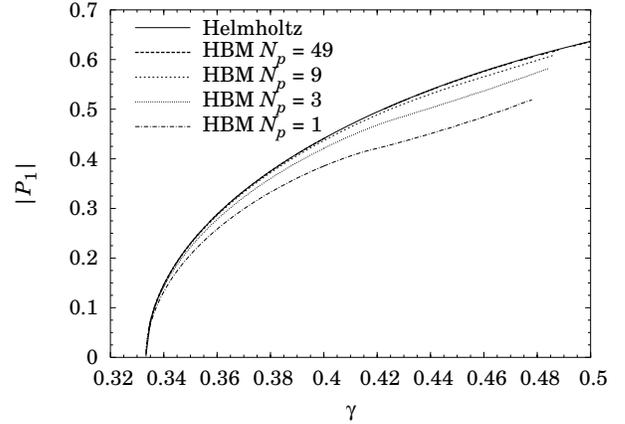}%
\else
  \ifoutputfig%
  \includegraphics[width=5.5in]{\psdir/HBM-hh-gamma.eps}%
  \fi
\fi
\caption
[Amplitude of first harmonic as the blowing pressure increases for
  the Helmholtz solution \eqref{e:helmotion} and the HBM truncated to
  1, 3, 9, and 49 harmonics, the last coinciding with Helmholtz
  ($\zeta=0.5$, $\eta=10^{-5}$)]
{\label{f:P1-gamma}
\ifgalleyfig
{Amplitude of first harmonic as the blowing pressure increases for
  the Helmholtz solution \eqref{e:helmotion} and the HBM truncated to
  1, 3, 9, and 49 harmonics, the last coinciding with Helmholtz
  ($\zeta=0.5$, $\eta=10^{-5}$)}
\fi}
\end{figure}

In Figure~\ref{f:P1-gamma} we have completed some of the curves that
we failed to make in Figure~\ref{f:holes}, and even increased the number
of harmonics, owing to the backtracking mechanism.  Admittedly, at
$N_p=49$, a few holes can still be seen, but the convergence is
significantly improved.

Here the amplitude of the first
harmonic is plotted for different numbers of harmonics as a function
of the blowing pressure $\gamma$ together with first harmonic of the
Helmholtz solution, deduced from equation~\eqref{e:helmotion}.  In
practice, the solution at $\gamma = 0.4$ was found and then
\emph{hbmap} was used to make Harmbal calculate solution for each of a large
number of subsequent values of $\gamma$ down to the oscillation threshold
by using the previous solution as initial value. The procedure was
repeated from $\gamma=0.4$ up to the point where the reed started to
beat, i.e.\ for $p<\gamma-1$ in equation~\eqref{e:simplenonlin}.  Without losses
(Helmholtz solution) the beating threshold does not arrive
before $\gamma=0.5$, and this should be expected for the nearly
lossless case studied with the HBM also.  However, the number of harmonics
$N_p$ taken into account in the HBM calculations is too small to
follow the sharp edges of the square signal.  The resulting overshoots
in $p(t)$, as seen in Figure~\ref{f:nearthres-nl}b, cause $p$ to
prematurely exceed the criterion for beating.  The beating threshold
converges to 0.5 as $N_p$ increases (see also Ref.\ \onlinecite{fritz04}). Note
that, for the
chosen value of  $\zeta$, it can be calculated following
Hirschberg \cite[eq.(45)]{hirschberg95} that above $\gamma \simeq
0.45$, the Helmholtz solution loses its
stability through a 
subharmonic bifurcation (a period-doubling occuring).

By Figure~\ref{f:P1-gamma} we can also verify that the model
experiences a direct Hopf bifurcation (which is known since the work
of Grand et al.\ \cite{grand97}). Thus, a single harmonic is enough to
study the solution around the threshold. Far from the threshold, more
harmonics have to be taken into account for $P_1$ to converge toward the
Helmholtz solution. This is not obvious and  for example
contradictory with the hypothesis made for the VTM
\cite{kergomard00}. Thus Harmbal 
appears as an interesting tool to evaluate the relevance of
approximate methods according to the parameter values.

\subsubsection{Helmholtz oscillation for a stepped conical tube}

The saxophone works similarly to the clarinet, but the bore has a
conic form.  In this section we compare the HBM calculations with 
analytical results, and in order to calculate the Helmholtz motion
when losses are ignored, we need to simplify the cone by assuming that
it consists of a sequence of $N$ sylinders of length $l$ and cross section
$S_i=\frac12i(i+1)S_1$, $S_1=S$ being the cross section of the smallest
cylinder, and $i=1,\dots,N$ (see Ref.\ \onlinecite{dalmont00}).  The
total length of the instrument is thus $L=Nl$.  The input impedance of
such a {\em stepped cone\/} may be written as
\begin{equation}
  Z_r(\omega)=\frac{2i}
    {\cot\!\left(\!\frac{\omega'}{4} - i\alpha(\omega')\right) 
   + \cot\!\left(\!\frac{\omega'}{4N}- i\alpha(\frac{\omega'}N)\right)},
\label{e:coneimp}
\end{equation}
where $\omega'\triangleq2\omega/(N+1)$ when $\omega=2\pi\dimvar
f/f_r$, where $f_r$ is the first eigenfrequency of 
this resonator.  We have ignored the dispersion term
here. Equation~\eqref{e:coneimp} is used instead of 
equation~\eqref{e:freqresponse}, and the damping 
$\alpha(\omega)=\psi\eta\sqrt{\omega/2\pi}$ is zero in the analytic
Helmholtz case and very small ($\eta=2\cdot10^{-5}$ below which
convergence became difficult) for the calculations with the HBM.

As before, the pressure amplitude of the ideal lossless case is
calculated by solving $u(p)=u(-Np)$, and two solutions are
possible:\footnote{This result corrects equation~(14) in
ref.~\onlinecite{dalmont00}} 

\begin{equation}
\begin{array}{rl}
  \llap{$p^{\pm}$}(\gamma)\!\!\!\!&=\displaystyle\frac{(N{-}1)(2{-}3\gamma)}{2(N^2-N+1)}\\
 &\displaystyle\pm\,  \frac{  \sqrt{(N{-}1)^2+(N{+}1)^2(-3\gamma^2{+}4\gamma{-}1)}}{2(N^2-N+1)}
\end{array}
\label{e:coneN}
\end{equation}
as long as $\gamma<1/(N+1)$ for the standard Helmholtz motion ($p^+$) and
$\gamma<N/(N+1)$ for the inverted one ($p^-$), which is unstable.
Above these limits $p^+=\gamma$ and $p^-=-\gamma/N$.  The magnitude of
the first harmonic of a square or rectangular wave is then given by 
\begin{equation}
  P_1^{\pm}(\gamma)=\frac{\sin\frac\pi{N+1}}{\frac\pi{N+1}}p^{\pm}(\gamma).
\label{e:P1p}
\end{equation}
For $N=1$, equation~\eqref{e:coneN} reduces to
equation~\eqref{e:helmotion}. For higher $N$, the pressure oscillation
becomes asymmetric.  

\begin{figure}
\ifgalleyfig%
  \includegraphics[width=3.25in]{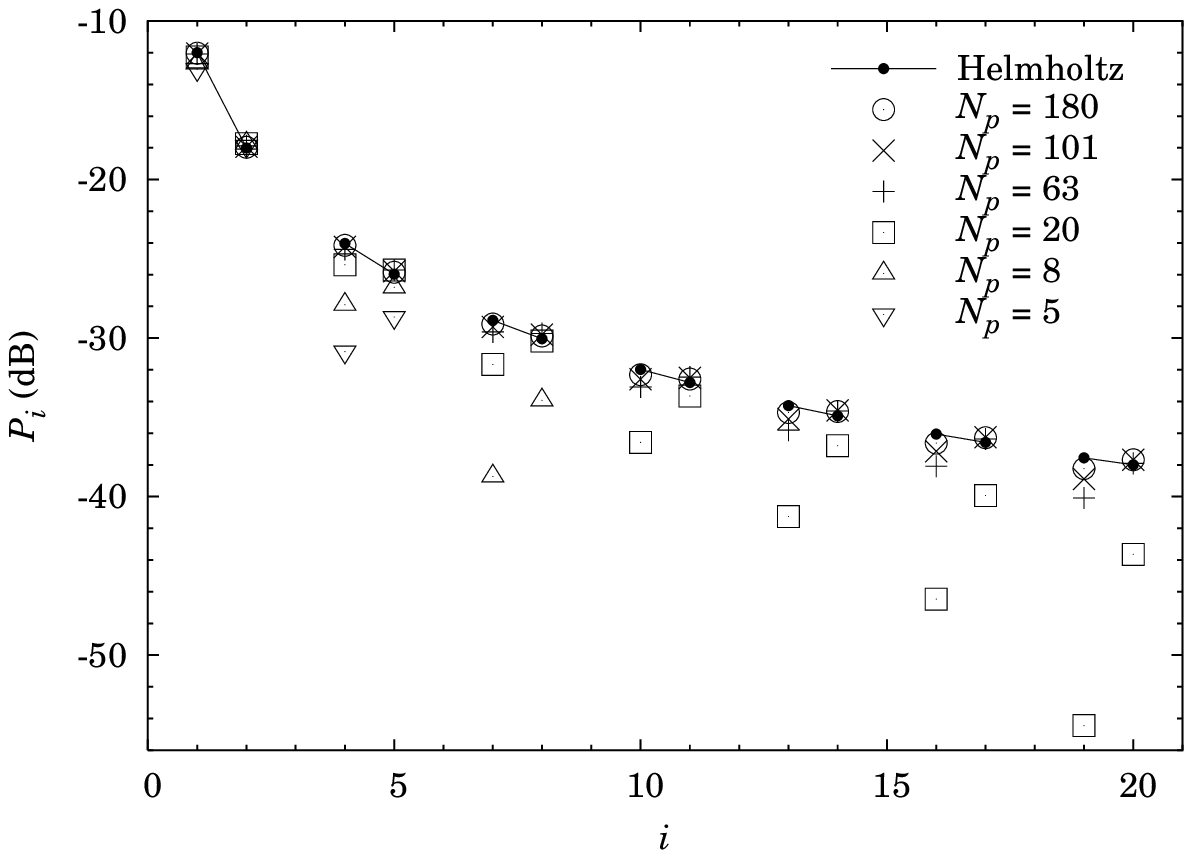}%
  \\\includegraphics[width=3.25in]{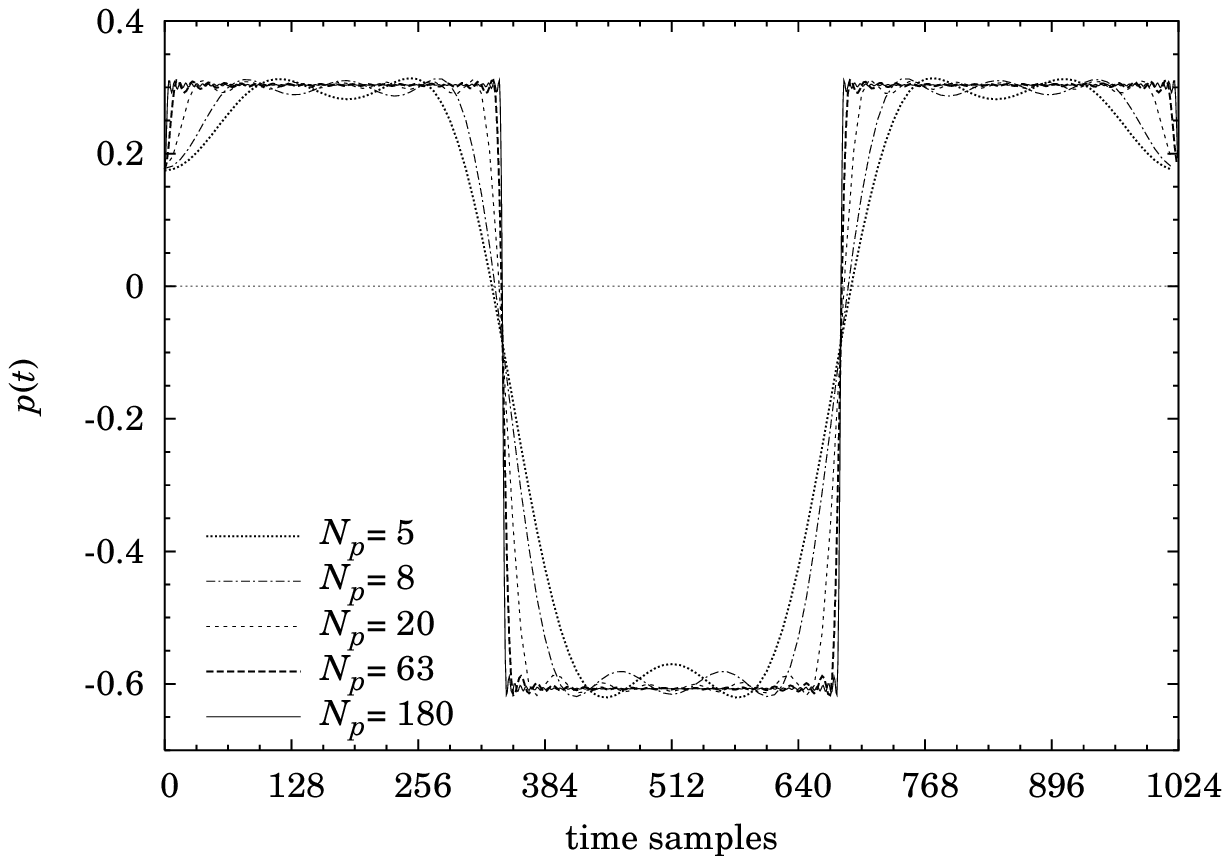}%
\else
  \ifoutputfig%
  \includegraphics[width=5.5in]{\psdir/freq-n2e-5g0.31A.eps}%
  \\\includegraphics[width=5.5in]{\psdir/time-n2e-5g0.31A.eps}%
  \fi
\fi
\caption
[Comparison between the standartd Helmholtz motion of a stepped cone
($N=2$) and the HBM for various $N_p$ at $\gamma=0.31$, $\zeta=0.2$, and
  $\eta=2\cdot10^{-5}$. (a) The magnitude of the harmonics and (b) one
  oscillation period. $N_t$ varies from 128 for $N_p{=}5$ to
  1024 for $N_p{=}180$.]
{\label{f:coneN2A}
\ifgalleyfig
{Comparison between the standartd Helmholtz motion of a stepped cone
($N=2$) and the HBM for various $N_p$ at $\gamma=0.31$, $\zeta=0.2$, and
  $\eta=2\cdot10^{-5}$. (a) The magnitude of the harmonics and (b) one
  oscillation period. $N_t$ varies from 128 for $N_p{=}5$ to
  1024 for $N_p{=}180$.}
\fi}
\end{figure}

We take the case $N=2$ and get 
\begin{equation}
  p^{\pm}(\gamma)=\frac16\left(2-3\gamma\pm\sqrt{-27\gamma^2+36\gamma-8}\right).
\label{e:coneN2}
\end{equation}
This result is compared with HBM calculations in
Figure~\ref{f:coneN2A} for $\gamma=0.31$.
Theoretically, the spectrum of the Helmholtz solution,
Figure~\ref{f:coneN2A}a, shows that every third component is missing
(actually zero) while the remaining components decrease in
magnitude thus 
forming the asymmetric pressure oscillation as shown in
Figure~\ref{f:coneN2A}a.  The HBM, on the other hand, suggests that
the first component in each pair be smaller than the second component.
This results in a {\em dip\/} at the middle of the long, positive part
of the period (i.e. on both extremities $t=0$ and $t=1024$ of the
curve in Figure \ref{f:coneN2A}).  The same was observed for $N=3$ and $N=4$, where the long
part of the period was divided by similar dips into three and four parts,
respectively (not shown).  The number of
time samples, $N_t$ did not change this fact, but as
Figure~\ref{f:coneN2A} indicates, the dips gradually become narrower
as the number of harmonics $N_p$ increases.  This indicates that the
HBM approaches the Helmholtz solution as $N_p$ approaches infinity.
\begin{figure}
\ifgalleyfig%
  \includegraphics[width=3.25in]{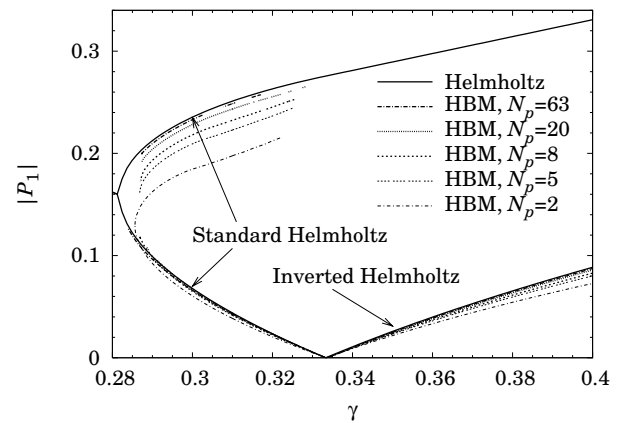}%
\else
  \ifoutputfig%
  \includegraphics[width=5.5in]{\psdir/map-n2e-5P1.eps}%
  \fi
\fi
\caption
[Amplitude of first harmonic $P_1$ as a function
of the blowing pressure $\gamma$ for
  the Helmholtz solution \eqref{e:coneN2} for 2-stepped cone and the
  HBM truncated to 2, 5,\dots, and 63 harmonics, the last coinciding
  with Helmholtz ($\zeta=0.2$, $\eta=2\cdot10^{-5}$). Only nonbeating
  regimes are shown.]
{\label{f:P1-gammaN2}
\ifgalleyfig
{Amplitude of first harmonic $P_1$ as a function
of the blowing pressure $\gamma$ for
  the Helmholtz solution \eqref{e:coneN2} for 2-stepped cone and the
  HBM truncated to 2, 5,\dots, and 63 harmonics, the last coinciding
  with Helmholtz ($\zeta=0.2$, $\eta=2\cdot10^{-5}$). Only nonbeating
  regimes are shown.}
\fi}
\end{figure}

A bifurcation diagram is plotted in Figure~\ref{f:P1-gammaN2}.
Similarly to Figure~\ref{f:P1-gamma} for the cylindrical bore, the
amplitude of the first harmonic is plotted for different number of
harmonics as a function of the blowing pressure $\gamma$. The Helmholtz
solution (equation~\eqref{e:P1p} with $N=2$) is also plotted.
As shown by Ollivier et al.\ \cite{ollivier04b}, the lower part of the upper branch and the branch of the inverted
Helmholtz motion are unstable.

In practice, these curves are more difficult to obtain with 
{\em hbmap\/} than for the cylindrical bore, especially close to the
subcritical oscillation threshold around $\gamma=0.28$, where computation was not
possible at this low losses.  More sophisticated continuation schemes should be
considered to obtain complete curves.  However, it is obvious from the
diagram that the model experiences a sub-critical Hopf bifurcation,
which agrees with the conclusion of Grand et al.\ \cite{grand97}.  This means that a
single-harmonic approximation is not enough to study the solution
around this threshold, since the small-amplitude hypothesis does not
hold. Further from the threshold, convergence toward the Helmholtz
motion is ensured as the number of harmonics $N_p$ is increased.

Only the nonbeating reed regime is considered in the figure and,
similarly to Figure~\ref{f:P1-gamma}, it can be noted that the beating
threshold for the model with $N_p$ harmonics depends on $N_p$ but
converges toward the Helmholtz threshold $\gamma=1/3$ (corresponding
to the lossless, continuous system) as $N_p$ is increased.

\subsubsection{Validation with time-domain model}

When adding a mass and damping to the reed or viscous losses and
dispersion to the pipe, it is more difficult to compare Harmbal
results with analytic solutions. This has been done by Fritz et al.\
\cite{fritz04} 
as far as the playing frequency is concerned, by comparison with
approximate analytical formula. Here, we propose to confront both the
playing frequency and the amplitude of the first partial with numerical results obtained with a time-domain method. We use a
newly developed (real-time) time-domain method (here called TDM) by
Guillemain et al.\ \cite{guillemain03a}.  It is based on the same set
of equations as presented in Section~\ref{s:clarinet} except that the
impedance of the bore is slightly modified to be expressed as an
infinite impulse response.  In the Fourier domain, it can be expressed
as
\begin{equation}
 Z_r(\althat\omega)=\frac{1-a_1e^{-i\althat\omega}-b_0e^{-i\althat\omega D}} 
       {1 - a_1e^{-i\althat\omega} + b_0 e^{-i\althat\omega D}}.
\label{e:philimp}
\end{equation}
where $\althat\omega=\dimvar\omega/f_s$, $f_s$ being the sampling
frequency, and the integer $D=\mathrm{round}(f_s/2f_r)$ the time delay
in samples for the sound 
wave to propagate to the end of the bore and back.
The constants $a_1$ and $b_0$ are to be adjusted so that the two
first peaks of resonance have the same
amplitude as the two first peaks of 
equation~\eqref{e:freqresponse}.

To express equation~\eqref{e:philimp} using our terminology,
we remember that $\omega=2\pi\dimvar f/f_r$ and obtain
\begin{equation}
 Z_r(\omega)=\frac{1 - a_1e^{-i\omega\frac{f_r}{f_s}} - b_0 e^{-i\omega/2}}
             {1 - a_1e^{-i\omega\frac{f_r}{f_s}} + b_0 e^{-i\omega/2}}.
\label{e:philimp2}
\end{equation}

\begin{figure}
\ifgalleyfig%
  \includegraphics[width=3.25in]{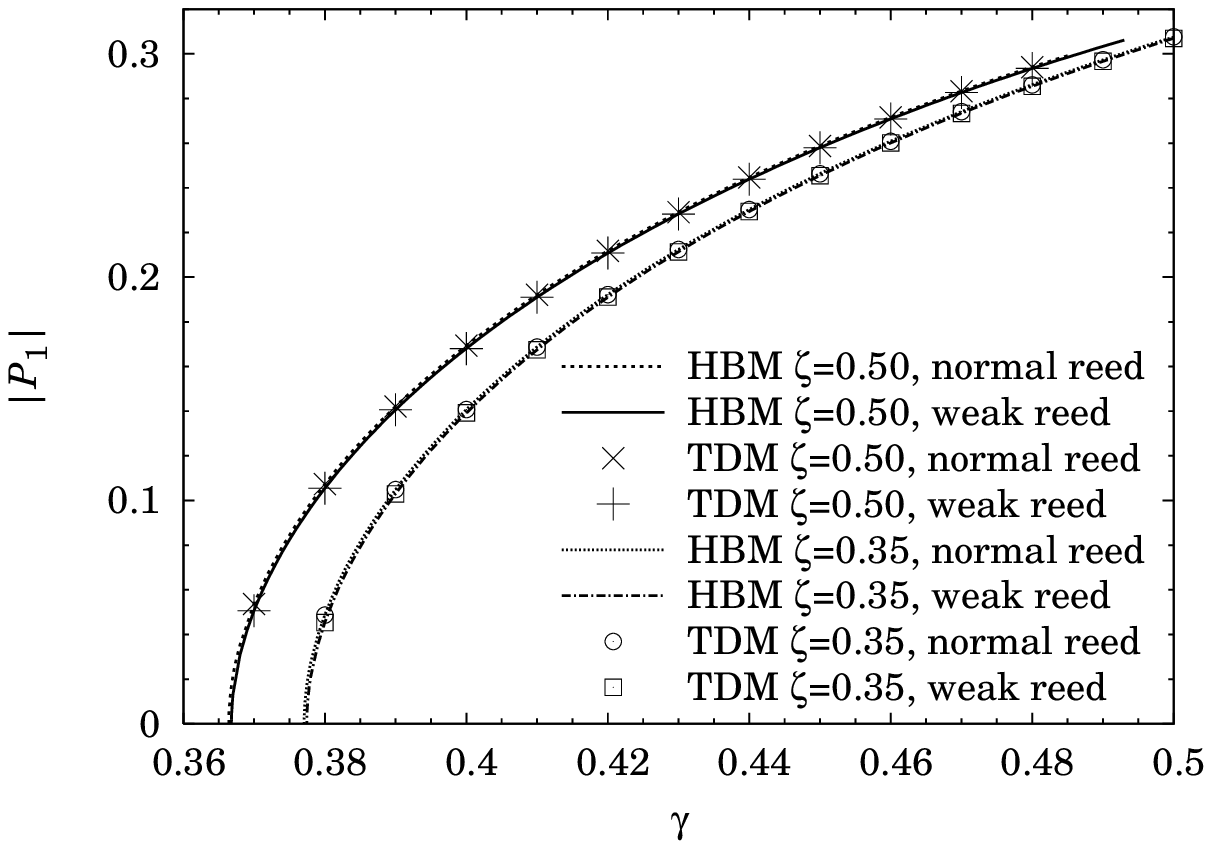}%
  \\\includegraphics[width=3.25in]{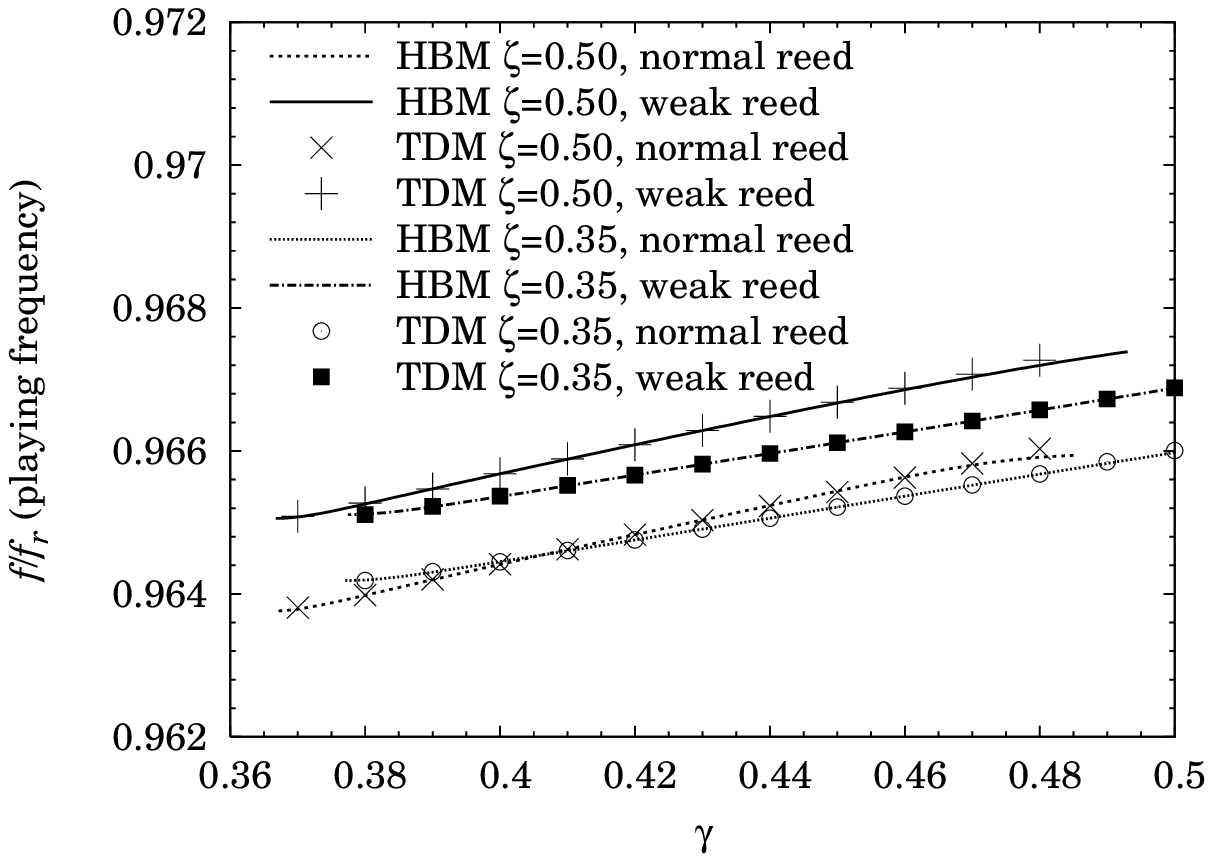}%
\else
  \ifoutputfig%
  \includegraphics[width=5.5in]{\psdir/phil-gamma-p15.eps}%
  \\\includegraphics[width=5.5in]{\psdir/freqcomp.eps}%
  \fi
\fi
\caption
[Comparison between HBM and TDM of the amplitude of (a) the
  first harmonic $P_1$ and (b) the dimensionless playing frequency
  $f_p/f_r$ as the blowing pressure $\gamma$ increases for a
  clarinet-like system with viscous losses and weak and normal reed
  interaction.  TDM values for $\zeta=0.50$ and $\gamma>0.48$ are
  omitted due to period doubling. So are the beating regimes of HBM
  calculations. 
  ($f_s\,{=}\,51100$Hz, $N_t\,{=}\,512$, $f_r\,=\,103.4$\,Hz, $N_p\,{=}\,15$)]
{\label{f:phil-gamma}
\ifgalleyfig
{Comparison between HBM and TDM of the amplitude of (a) the
  first harmonic $P_1$ and (b) the dimensionless playing frequency
  $f_p/f_r$ as the blowing pressure $\gamma$ increases for a
  clarinet-like system with viscous losses and weak and normal reed
  interaction.  TDM values for $\zeta=0.50$ and $\gamma>0.48$ are
  omitted due to period doubling. So are the beating regimes of HBM
  calculations. 
  ($f_s\,{=}\,51100$Hz, $N_t\,{=}\,512$, $f_r\,=\,103.4$\,Hz, $N_p\,{=}\,15$)}
\fi}
\end{figure}

In this section, we also include the mass and damping of the reed,
so $M$ and $R$ are no longer zero.  The TDM does not work for
$M=R=0$, or even for values close to this, so we have used a reed with
weak interaction with the pipe resonance as well as one with close to
normal reed impedance.  The corresponding values for $\omega_e$ and
$q_e\triangleq g_e/\omega_e$ are shown in Table~\ref{t:phil}.
\begin{table}
  \caption{The values of $M$ and $R$ for three strengths of reed
  interaction. The bore parameters are $D=247$ ($f_r=103.4$\,Hz),
  $a_1=0.899$, and $b_0=0.0946$ for sampling frequency
  $f_s=51100$\,Hz.}
  \smallskip
  \centerline{%
    \begin{tabular}{l|cccc}\hline
      Reed &$\omega_e$/Hz&$q_e$ &$M$&$R$\\\hline
      Weak   &10000        & 0.1  &$1.070\e{-4}$ &$1.034\e{-3\vphantom{^1}}$\\
      Normal &\ph02500     & 0.2  &$1.712\e{-3}$ &$\ph08.28\e{-3}$\\
      \hline
    \end{tabular}}
  \label{t:phil}
\end{table}
Figure~\ref{f:phil-gamma}a shows the bifurcation diagram for two values
of $\zeta$ and for weak and normal reed impedance, 
while Figure~\ref{f:phil-gamma}b shows the corresponding variation in
the dimensionless playing frequency $f_p/f_r$.  The lines represent the
continuous solutions of the HBM, and the symbols show a set of results
derived from the steady-state part of the TDM signal.
The TDM symbols fall well on the lines of the HBM, except for
$\zeta=0.50$ when $\gamma$ approaches 0.5.  Then the TDM experiences
period doubling, i.e.\ two subsequent periods of the signal differ.
At the same time, not being able to show subharmonics, the HBM shows
signs of a beating reed, possibly a solution that is unstable and thus
not attainable by time-domain methods. 

Note that three points have to be verified before comparing results
from the HBM and the TDM:

The numerical scheme used in the TDM to approximate the time
derivatives in the reed equation~\eqref{e:lindiff} requires
discretization.  Depending on the sampling 
frequency $f_s$, the peak of resonance of the reed  deviates more
or less from the one given by the continuous equation.  For normal
reed interaction 
($f_e$=2500 Hz), the deviation is negligible, but it may
become significant in the case of weak reed interaction, where the peak is at
10000\,Hz. However, the 
fact that the reed and the bore interact weakly in the latter case, implies
that the exact position of the peak has little importance.  Therefore,
at the used sampling frequency, the discretization in the TDM is not
compensated for in the HBM calculations.

Then there should be agreement between the sampling frequency $f_s$ in the
TDM and the number of samples $N_t$ per period in the HBM. Their
relation is given by
\begin{equation}
N_t=\frac{f_s}{f_p}.
\end{equation} 
In order to have a sufficiently high sampling rate, we have chosen
$N_t=512$. The playing frequency $f_p$ is plotted in
Figure~\ref{f:phil-gamma}b, and we used an average
$f_s=51100$\,Hz for both the HBM and the TDM.

Finally, it seems also necessary that $N_p$ and $N_t$ are chosen so that
\begin{equation}
N_p+1 = \frac{N_t}{2}.
\end{equation}
In practice, however, when comparing bifurcation diagrams of the first
harmonic $P_1$, as in Figure \ref{f:phil-gamma}, rather low values of
$N_p$ give good results. Nevertheless, more harmonics are obviously
needed to compare waveforms in the time domain, especially far from the
oscillation threshold.

\section{Practical experiences}\label{s:practexp}\label{s:disc}

\subsection{Multiple solutions}

As we consider a nonlinear problem, we cannot anticipate the number of
solutions. Therefore, it should not be surprising that it is possible
to obtain multiple solutions for a given set of parameter values.
When searching for a particular solution, this may be a practical
problem. Fritz et al.\ \cite{fritz04} have discovered that some
solutions seem to disappear when increasing the number of harmonics
$N_p$, implying that solutions may arise from the truncation to a
finite $N_p$. We have now discovered alternative solutions that
persist even at very high $N_p$.

\begin{figure}
\ifgalleyfig%
  \includegraphics[width=3.25in]{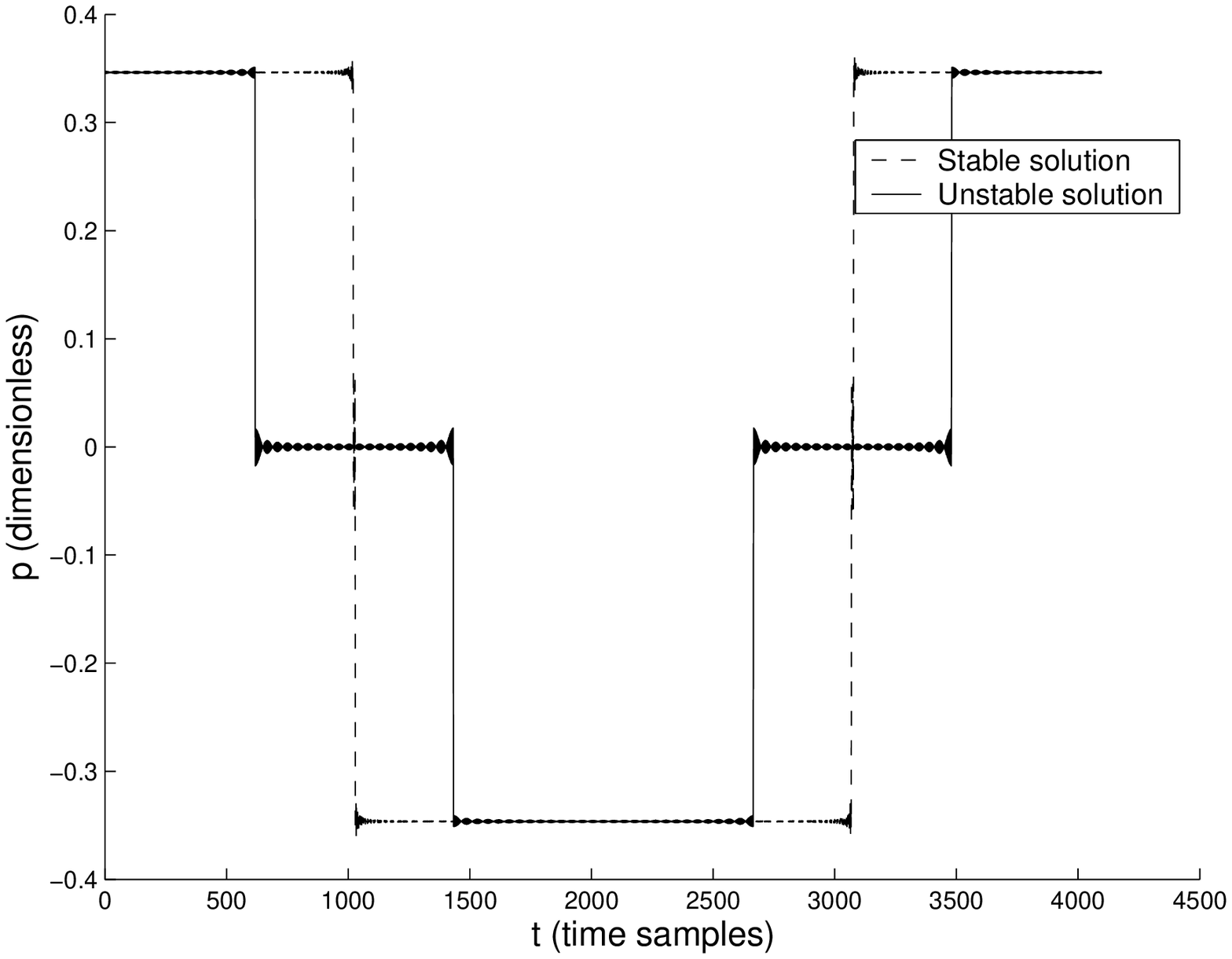}%
  \\\includegraphics[width=3.25in]{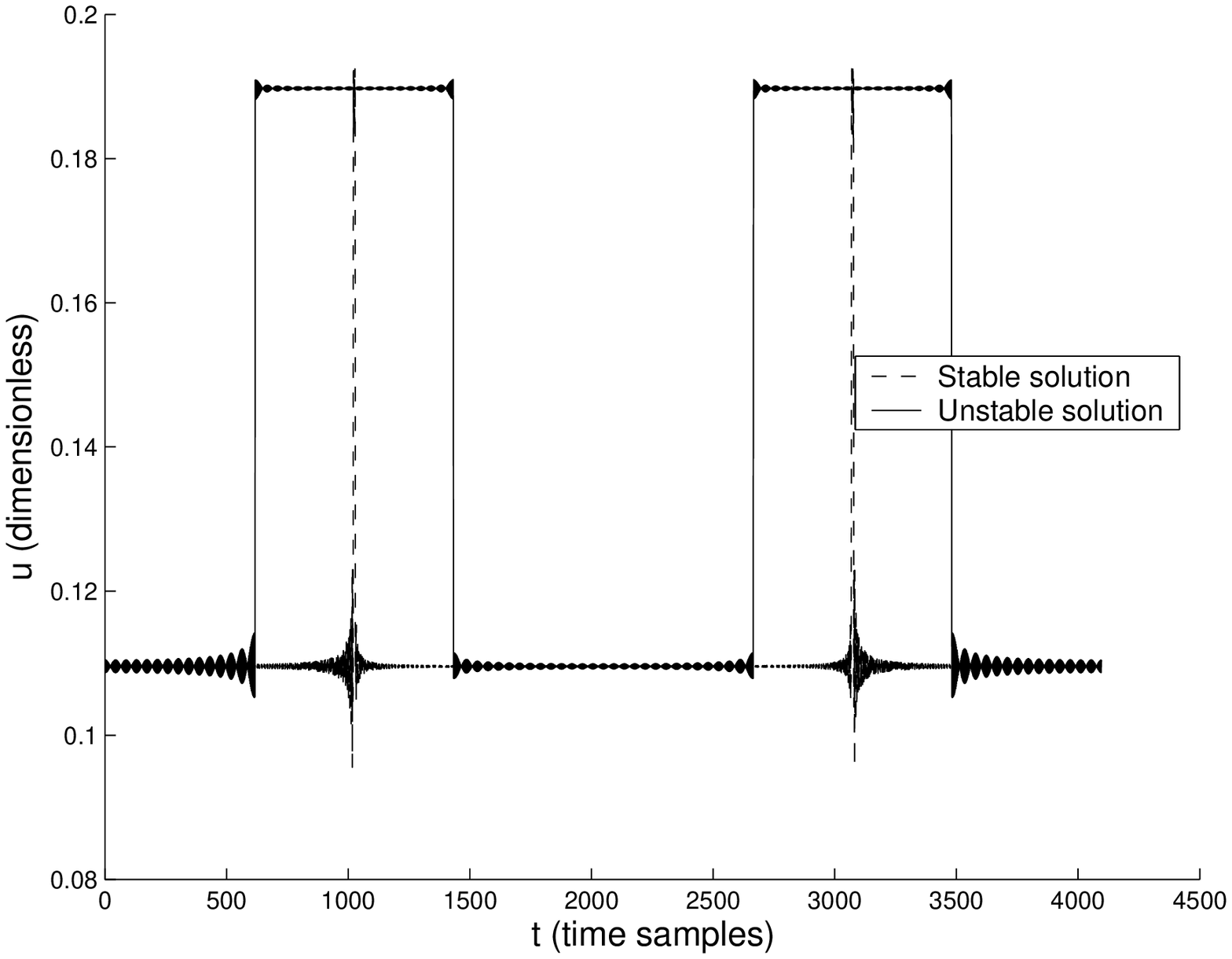}%
\else
  \ifoutputfig%
  \includegraphics[width=5.5in]{\psdir/comp_p_helmbiz2.eps}
  \\\includegraphics[width=5.5in]{\psdir/comp_u_helmbiz2.eps}
  \fi
\fi
\caption
[The pressure (a) and volume-flow (b) wave form of the
Helmholtz solution and a 3-level sister solution calculated by 
the HBM employing the simple clarinet model with $\zeta=0.5$, $\gamma=0.4$,
$N_p=2000$, $\eta=10^{-5}$.]
{\label{f:bizsol}
\ifgalleyfig
{The pressure (a) and volume-flow (b) wave form of the
Helmholtz solution and a 3-level sister solution calculated by 
the HBM employing the simple clarinet model with $\zeta=0.5$, $\gamma=0.4$,
$N_p=99$, $\eta=10^{-5}$.}
\fi}
\end{figure}

Let us illustrate this with the simple model of the clarinet used in
Section~\ref{s:helcyl}, where the reed is a spring without mass or
damping, the nonlinearity is given by equation~\eqref{e:simplenonlin},
and the bore is an ideal cylinder with nearly lossless propagation
and no dispersion.  Figure~\ref{f:bizsol} shows a three-level
sister solution together with the related Helmholtz solution for a
large number of harmonics, $N_p=2000$.

A solution of the lossless problem should satisfy the criteria
\cite{kergomard95}
\begin{equation}
\left\{
\begin{array}{l}
p(t+\pi)=-p(t)\\
u(t+\pi)=u(t)\\
\end{array}
\right.
\label{e:pu_temp}
\end{equation}
(the dimensionless period being $2\pi$),
as well as the conditions stated before equation~\eqref{e:helmotion},
noting that $p(t)\,{=}\,u(t)\,{=}\,0$ for all $t$ is the static solution. It
is easily verified graphically that both of the solutions in 
Figure~\ref{f:bizsol} satisfy these conditions. Moreover, since they
also satisfy equation~\eqref{e:simplenonlin}, the three-level
solution is a solution of the lossless model.

\begin{figure}
\ifgalleyfig%
  \includegraphics[width=3.25in]{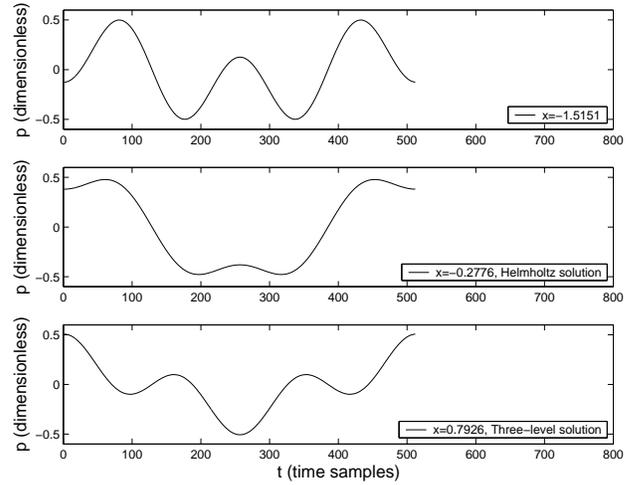}%
\else
  \ifoutputfig%
  \includegraphics[width=5.5in]{\psdir/helm_biz_3h_bis.eps}%
  \fi
\fi
\caption
[The pressure waveform of the
three solutions found by 
the HBM with $N_p=3$ employing the simple clarinet model with
$\zeta=0.5$, $\gamma=0.4$, $\eta=10^{-5}$.]
{\label{f:helm_biz_3h}
\ifgalleyfig
{The pressure waveform of the
three solutions found by 
the HBM with $N_p=3$ employing the simple clarinet model with
$\zeta=0.5$, $\gamma=0.4$, $\eta=10^{-5}$.}
\fi}
\end{figure}

\begin{figure}
\ifgalleyfig%
  \includegraphics[width=3.25in]{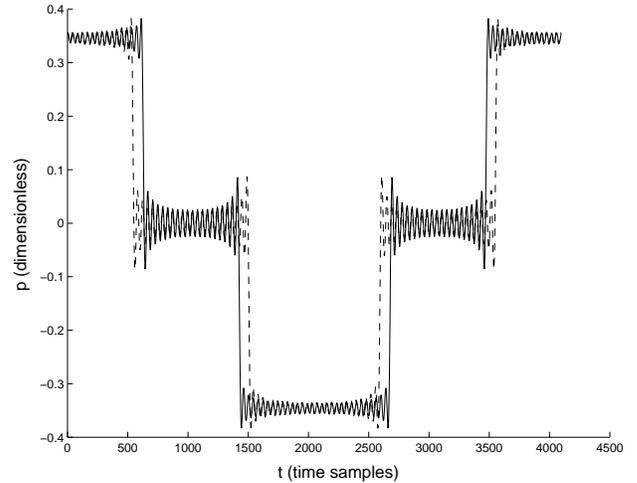}%
\else
  \ifoutputfig%
  \includegraphics[width=5.5in]{\psdir/deux_sol_np100.eps}%
  \fi
\fi
\caption
[The pressure waveform of two solutions that differ by the duration of their steps, found by 
the HBM employing the simple clarinet model with
$\zeta=0.5$, $\gamma=0.4$, $\eta=10^{-5}$, $N_p=99$.]
{\label{f:sol_np100}
\ifgalleyfig
{The pressure waveform of two solutions that differ by the duration of their steps, found by 
the HBM employing the simple clarinet model with
$\zeta=0.5$, $\gamma=0.4$, $\eta=10^{-5}$, $N_p=99$.}
\fi}
\end{figure}

Whereas the system of time-domain equations~\eqref{e:pu_temp} has an
infinity of solutions, truncation in frequency-domain limits the number
of solutions. The unique solution of the HBM with only one harmonic is
obviously a sine. Let us analyse the situation in the
simplest nontrivial case of the lossless problem with two odd
harmonics and a cubic 
expansion for nonlinear coupling. Ignoring even harmonics, the HBM gives a
system of two equations (see Kergomard et al.\ \cite{kergomard00}):
\begin{equation}
\left\{
\begin{array}{rclr}
        \alpha & = & 3P_12(1+x+2|x|^2) &
\mathrm{(a)}\\ 
        \alpha x & = &
P_12(1+3x|x|^2+6x), &
\mathrm{(b)}\\
\end{array}
\right.
\label{e:3h_P}
\end{equation}
where $\alpha=-A/C$ and $x=P_3/P_1$. As
equation~(\ref{e:3h_P}a) imposes $P_3$ to be 
real, solving this system amounts to solving
\begin{equation}
x^3+x^2-x=1/3.
\label{e:3h_x}
\end{equation}
This equation has three real solutions $x \simeq -1.5151$,
$-0.2776$ and $0.7926$. 
All of them are found by Harmbal for negligible losses ($\eta=10^{-5}$),
and the corresponding waveforms are presented in
Figure~\ref{f:helm_biz_3h}.  We note that the second solution leads to the Helmholtz motion
when increasing the number of harmonics (with the theoretical value
known to be $x=-1/3$) whereas the third one corresponds to the three-level solution in Figure \ref{f:bizsol}. We can also easily imagine that these three solutions of the truncated problem are three-harmonic approximations of square waves that are distributed on three levels: $p^\pm \simeq \pm 0.5$ and $p=0$. Respectively, they have two, zero, and one steps at the zero-level. 
It should be noted that the conditions (\ref{e:pu_temp}) for the continuous problem do not constrain the duration of each step. Figure \ref{f:sol_np100} shows  two such twin solutions for $N_p=99$ corresponding to the three-level solution in Figure \ref{f:helm_biz_3h}. This has to be kept in mind when increasing $N_p$ using the HBM.
 
While the Helmholtz motion is known to be stable \cite{kergomard95},
the two three-level solutions can be considered as a combination of the
static solution (the zero level) and the square wave (two levels with
opposite values). Since we know from Kergomard \cite{kergomard95} that
in the case of ideal propagation (neither losses nor dispersion), the
stability domain of these two solutions are mutually exclusive, it can
be concluded that the three-level solutions are unstable.

Taking into account losses in the propagation does not make the
three-level solutions vanish. But a
simple reasoning to determine the stability of this 
solution is  not possible in this case. To
the authors knowledge, 
however, such a solution has never been observed experimentally at low
level of excitation. 

\subsection{Initial value of the playing frequency}

A practical difficulty encountered is the convergence of the playing
frequency $f_p$.  If its initial value is not close enough to the
solution, divergence is almost inevitable.  
This occurs because the resonator impedance $Z_r$ tends to vanish
outside the immediate surroundings of the resonance peaks of the resonator, 
rendering $\vec F(\vec P,f_p)$ very small and thereby $\vec G\simeq
\vec P/P_1$ nearly constant with respect to $f_p$.  The slope
$\partial\vec G/\partial f_p$ thus becomes close to
zero, the Newton step leads far away from the solution, and
convergence fails.  Dissipation widens the resonance peaks and thus
also the convergence range.

For a simple system where the playing frequency is known to correspond
to a resonance peak of the tube, initializing $f_p$ is easy.  However,
with dispersion or other inharmonic effects, choosing an initial value
for $f_p$ may be difficult.  In Harmbal the problem may to some extent
be avoided by the possibility of gradually adding the
dispersion (or other inharmonic effects), so that the playing frequency can be
followed quasi-continuously from a known solution without dispersion,
for instance by using {\em hbmap}.

\section{Conclusions}\label{s:concl}

The harmonic balance method (HBM) is suited for studies of
self-sustained oscillations of musical instruments, and the
computer program {\em Harmbal\/} has been developed for this application.
It is available with its source
code \cite{harmbal}, has a free licence, and is already in use by
several researchers.  It is programmed in C, runs fast, and is easily
used by other application, such as for continuation purposes. 

Some difficulties are related to  the digital sampling of the signal
and can be solved by introducing a backtracking mechanism. When using
a large number of harmonics, the extreme case of the (lossless)
Helmholtz motion can be solved for different shapes of
resonators. Nevertheless, the value of the first harmonic $P_1$
seems to  be well predicted by lower values of $N_p$, in particular
close to the threshold of a direct bifurcation. 
For the saxophone we used a stepped-cone bore and observed one or more
dips during the longest part of the period, depending on the number of
steps.  These dips approach pure impulses
as $N_p$ increases.  The number of samples $N_t$ in a period showed to
be insignificant for these dips. 

The HBM can lead to some alternative solutions for a unique
set of parameters. The nondissipative versions of these
solutions satisfy the continuous model 
equations, but  they are not stable and thus cannot be attained by ab
initio time-domain calculations.  Another
problem is the great sensitivity to the guessed playing frequency. 

As a consequence, a certain expertise is needed in order to use the
method, but, thanks to an automatic continuation procedure, the
calculation is easy. We note that also experimental
results can be used for the impedance of the resonator. 

\begin{acknowledgments}
The Europeen Union through the MOSART project is acknowledged for
financial support. We would also like to thank Claudia Fritz at IRCAM
in Paris for thorough testing and valuable feedback, Jo\"el Gilbert at
Laboratoire d'Acoustique de l'Universit\'e du Maine (LAUM) in Le Mans,
and Philippe Guillemain at the Laboratoire de M\'ecanique et
d'Acoustique at CNRS in Marseille for fruitful discussions during the
work, and the latter also for kindly providing some Matlab code for the
time-domain model.
\end{acknowledgments}

\bibliographystyle{jasasty}
\bibliography{fvkl-050218jasa}

\maketablesandfigures

\end{document}